\documentstyle[12pt,emulateapj]{article}

\begin{document}

\lefthead{S. Nitta}
\righthead{Transition from fast reconnection to slow reconnection}

\title{Continuous transition from fast magnetic reconnection to slow reconnection and change of the reconnection system structure}

\author{Shin-ya Nitta \altaffilmark{1}, \altaffilmark{2}}
\affil{ Solar-B Project Office, National Astronomical Observatory of Japan, 2-21-1 Osawa, Mitaka, Tokyo, 181-8588, Japan}
\authoremail{nittasn@yahoo.co.jp}

\altaffiltext{1}{ Department of Information and Communication Engineering, 
The University of Electro-Communications,  
1-5-1 Chofu-ga-oka, Chofu, Tokyo, 182-8585 Japan}

\altaffiltext{2}{Hayama Center for Advanced Studies, 
The Graduate University for Advanced Studies (SOKENDAI), 
Shonan Village, Hayama, Kanagawa, 240-0193 Japan }

\begin{abstract}

This paper analytically investigates a series of two-dimensional MHD reconnection solutions over a wide variation of magnetic Reynolds number ($R_{em}^*$). A new series of solutions explains a continuous transition from Petschek-like fast regime to a Sweet-Parker-like slow regime. The inflow region is obtained from a Grad-Shafranov analysis used by Nitta et al. 2002 and the outflow region from a shock-tube approximation used by Nitta 2004, 2006. A single X-point (Petschek-like) solution forms for a sufficiently small $R_{em}^*$. As $R_{em}^*$ gradually increases, the solutions shifts to an X-O-X solution with a magnetic island between two X-points.  When $R_{em}^*$ increases further, the island collapses to a new elongated current sheet with Y-points at both ends (Sweet-Parker-like). These reconnection structures expand self-similarly as time proceeds. As $R_{em}^*$ increases, the reconnection rate and the reducible fraction of the initial magnetic energy of the system decrease as power-law functions of $R_{em}^*$.
\end{abstract}


\section{Introduction}
\label{sec:intro}
Majority believes that the magnetic reconnection is an effective and universal energy conversion process in many plasma systems. Especially, the magnetic reconnection plays essential roles in astrophysical phenomena, e.g., solar flares (see Tsuneta 1996), geomagnetospheric substorms, explosive phenomena in YSOs (see Koyama 1994, Hayashi et al. 1999). 

To apply the reconnection theory to astrophysical phenomena, the particular condition of the astrophysical reconnection must be considered: It is intrinsically time-dependent and works in a ``free-space'' in which any environmental condition does not influence the evolution of the reconnection system. In such condition, the self-similar evolution model (see Nitta et al. 2001; hereafter paper 1) of the reconnection will be relevant. In this paper, I discuss the dependence of the self-similar reconnection model on the wide variation of the electric resistivity. 

The entire scale of reconnection system is much larger than typical particle scales, e.g., the ion Larmor radius. This fact enables us to treat the reconnection problems in a regime of magnetohydrodynamics (MHD). The MHD approximation is very useful to understand the basic macroscopic physics of the magnetic reconnection. Many theoretical and numerical works based on MHD succeeded to clarify the nature of the MHD reconnection and proposed its applications (see Sweet 1958, Parker 1963, Petschek 1964, Vasyliunas 1975, Priest \& Forbes 1986, a series of papers by Ugai and his collaborators, e.g., Ugai \& Zheng 2005, Sato \& Hayashi 1979 etc.). The magnetic reconnection properly takes place in resistive plasmas. We should remember here that the electric resistivity cannot be reduced from MHD equations, but from some microscopic particle scale physics. This has been the essential problem in reconnection studies based on the framework of MHD.  

It is widely accepted that localized large electric resistivity enhanced in the middle of the current sheet is required in order to establish a high power (fast) reconnection (Biskamp 1986, Scholer 1989, Yokoyama \& Shitaba 1994). Baty et al. (2006) claim a somewhat different result; maintaining the Petschek-like fast reconnection does not require significantly localized large resistivity. However, even in numerical simulation by Baty et al. (2006), it initially needs a localized (but not so large) resistivity to set up a Petschek-like fast reconnection. These results may suggest that it needs a localized resistivity to determine the reconnection point in the elongated current sheet system at least in the onset stage. 

Thus we may state that, in MHD reconnection, the so-called anomalous resistivity is required to establish the fast reconnection. Because the origin of the anomalous resistivity is a microscopic process and outside the scope of MHD study, we must represent it by relevant model including parameter(s). Unfortunately we neither have identified the definite microscopic elementary process of the anomalous resistivity nor have understood its typical value yet (see Coppi \& Friedland 1971, Ji et al. 1998, Shinohara et al. 2001). The value of enhanced resistivity may depend upon several factors. Recent particle simulations for the LHD instability (one of the probable candidate for microscopic process of the resistivity enhancement) revealed that the resistivity crucially depends upon the parameters: the current sheet thickness, the ratio of the electron plasma frequency to the electron gyro frequency, the guide magnetic field strength (see, Tanaka et al. 2006). This result inspires us to study the influence of the variation of the electric resistivity even in the same object, e.g., solar flares. 

Hence, in this paper, I discuss the MHD reconnection in a free space for a very wide variation of the electric resistivity. The key parameter of this study is the magnetic Reynolds number $R_{em}^*$ defined in section \ref{sec:concept} in accordance with the magnitude of the electric resistivity. 

For the case of sufficiently small magnetic Reynolds number ($R_{em}^* \sim 15$), I discussed the plasma-$\beta$ dependence of the self-similar model in Nitta 2004; hereafter paper 3. As a result, I had clarified that the reconnection system is insensitive to the variation of plasma-$\beta$. The reconnection rate $R^*$ has rather universal value $R^* \sim 0.05$ for $\beta \ll 1$ and $R^* \sim 0.034$ for $\beta \gg 1$ if the resistivity is sufficiently large (see figure 2 of paper 3, more precise result is in figure 2 of Nitta 2006; hereafter paper 4).

Paper 4 discussed the magnetic Reynolds number dependence for the particular case that the reconnection point is fixed at a point in the middle of the current sheet. In this treatment, only the case with single X-point can take place (the Petschek-like solution). When the magnetic Reynolds number increases, a remarkable converging inflow concentrates into the region near the diffusion region. The direction of the converging inflow gradually tends to be parallel to the magnetic field lines as the magnetic Reynolds number increases. This converging inflow directly results in the reduction of the reconnection rate $R^*$ because it is proportional to the electric field (the vector product of the velocity and the magnetic field) around the diffusion region. Thus we obtained the result $R^* \propto {R_{em}^*}^{-1}$. However, the converging inflow becomes too strong for $R_{em}^* \mathrel{\hbox{\rlap{\hbox{\lower5pt\hbox{$\sim$}}}\hbox{$>$}}} 30$: the converging speed exceeds the Alfv\'{e}n speed $V_{A0}$ at the asymptotic region (note $V_{A0}$ is the maximum limit for spontaneous flow speed). This is an irrational result and showing the limit of validity of the treatment in paper 4. The cause is likely from the approximation that  the locus of the reconnection point is fixed. This deficiency must be improved to be applicable for the case of higher magnetic Reynolds number. 

The purpose of this paper is to clarify how large is the reconnection rate and how is the dependence on the reconnection structure at large magnetic Reynolds number region in a more general circumstance. The hint is in the result of paper 4: the converging inflow in a large magnetic Reynolds number makes a channel flow structure (see section 4.1 of paper 4). From the discussion of the channel flow structure, we can speculate the following new type of solutions. 

In the channel flow, the inflow speed toward the slow shock is much larger than that toward the diffusion region. This non-uniform inflow leads a non-uniform compression of the field reversal region. Since the compression outside the diffusion region is much stronger than that in the diffusion region, the reconnection point may move outside the present diffusion region. Thus, an O-point will form around the original diffusion region. Consequently, a typical X-O-X structure will form (see section 4.6 of paper 4). 

In order to check the above speculative idea for very large magnetic Reynolds number, I extend the shock tube approximation used in papers 3 and 4 to more general case with moving X-point. The alterations from the previous works are listed in section \ref{sec:difference}. 

This paper is organized as follows. The self-similar reconnection model is briefly summarized in section \ref{sec:ss-model}. The numerical scheme to solve the shock-tube approximated model is shown in section \ref{sec:num-sch}. The main result is listed in section \ref{sec:res}. The detailed meaning of the obtained results and its applications are discussed in section \ref{sec:sum-dis}.

\section{Self-similar reconnection model}
\label{sec:ss-model}
We must note that many cases of the actual magnetic reconnections in astrophysical systems usually grow over a huge dynamic range in their spatial dimension. For example, the initial scale of the reconnection system can be defined by the initial current sheet thickness, but this is too small to be observed in typical solar flares. We do not have any convincing estimate of the scale, but if we estimate it to be of the order of the ion Larmor radius, it is extremely small ($\sim 10^0$ [m] in the solar corona). Finally, the reconnection system develops to a scale of the order of the initial curvature radius of the magnetic field lines ($\sim 10^7$ [m] $\sim$ 1.5\% of the solar radius for typical solar flares). The dynamic range of the spatial scale is obviously huge ($\sim 10^7$ for solar flares). For geomagnetospheric substorms, their dynamic range of growth is also large ($\sim 10^4$ for substorms). 

In addition, we must note that, in many cases, the environmental timescale of energy storage (the convection timescale; typically of the order of day $\sim 10^5$ [sec] for solar active regions) is much longer than the evolution timescale of the reconnection (the Alfv\'{e}n transit timescale; typically of the order of $10^2$ [sec] for solar flares). This is common with the geomagnetospheric substorms. 

Above mentioned properties (very wide dynamic range of the expansion and the discrepancy of the timescales between the external circumstance and the reconnection system) suggest that the external circumstance is almost stationary during the evolution of the reconnection system. This means that the evolution of the magnetic reconnection should be treated as a spontaneous development in a ``free space'', and that the external circumstance does not affect the evolutionary process of magnetic reconnection, at least at the expanding stage just after the onset of reconnection. We must study the spontaneous development of the magnetic reconnection for the astrophysical applications. 

Let us suppose the initial state to be a two-dimensional equilibrium state with anti-parallel magnetic field distribution, as in the Harris solution. When non-uniform magnetic diffusion takes place in the current sheet by some localized resistivity, magnetic reconnection will start, and a pair of reconnection jets is ejected along the current sheet (see section 2 of paper 1). This causes a decrease in total pressure around the reconnection point. Such information propagates outward as a rarefaction wave. In a low-$\beta$ plasma ($\beta \ll 1$ in the region very distant from the current sheet [asymptotic region]; as typically encountered in astrophysical problems), the propagation speed of the fast-magnetosonic wave is isotropic, and is much faster than the slow mode wave. Thus, information about the decreasing total pressure propagates almost isotropically as a fast-mode rarefaction wave (hereafter FRW) with a speed almost equal to the Alfv\'{e}n speed $V_{A0}$ in the asymptotic region. Hence, the wave front of the FRW (hereafter FRWF) has a cylindrical shape except near the point where the FRWF intersects with the current sheet. The region swept by FRWF induces the inflow toward the reconnection region owing to the total-pressure difference by the FRW. When the FRWF sufficiently expands, the initial thickness $D_0$ of the current sheet becomes negligible comparing with the system size $V_{A0} t$, where $t$ is the time from the onset of reconnection ($V_{A0} t \gg D_0$). In such a case, there is only one characteristic scale, i.e., the radius of the FRWF ($V_{A0} t$), which linearly increases as time proceeds. This is just the condition for self-similar growth. 

When the inflow toward the current sheet develops sufficiently, a pair of slow shocks forms along the current sheet (the Petschek-like slow shock). After the formation of the slow shock, the energy conversion drastically proceeds. The most of the magnetic energy is converted on the slow shock. Hence this stage is a ``fast reconnection''. Once this system of fast reconnection is set up, the dimension of the system unlimitedly develops self-similarly (see figures 5-7 of paper 1) until the FRWF reaches the external spatial scale.

\section{Numerical Scheme}
\label{sec:num-sch}

\subsection{Concept}
\label{sec:concept}
The basic concept of the shock-tube approximation is the same with that of paper 4. As we mentioned in paper 3, the reconnection outflow in our spontaneous model has a structure involving several discontinuities like a shock-tube problem (see figure \ref{fig:structure-sch} ). We solve the reconnection outflow region by approximating as a kind of shock-tube problem (see details in the following subsections). In paper 4, we assumed a converging inflow toward the slow shock, and discussed the structure of the outflow region as a function of the magnetic Reynolds number. Consequently, we obtained the magnetic Reynolds number dependence of the structure (see figure 7 of paper 4) and the reconnection rate (see figure 6 of paper 4). 

In this paper we study the magnetic Reynolds number dependence of the outflow structure by the above shock-tube approximation in the regime of ideal (non-resistive) MHD in a more universal situation than that in paper 4. 

One may think that it is very curious that we can treat the magnetic Reynolds number in the regime of ideal MHD in this work. We should note, however, that if we can evaluate the reconnection electric field $E_z$ (note that $E_z$ is uniform in the vicinity of the reconnection point from the Ampere's law because this region is almost stationary in our self-similar model) and the magnetic field $B_x^*$ around the diffusion region from the shock-tube problem, we can estimate the inflow speed $v_y^*\ (\equiv E_z/B_x^*)$ toward the diffusion region. This inflow speed is in balance with the magnetic diffusion speed $v_{dif}^*$ at the diffusion region. Providing that the Alfv\'{e}n speed $V_{A0}$ in the asymptotic region to be fixed as a normalization value as in paper 3, inverse of the inflow speed toward the diffusion region shows the ``effective'' magnetic Reynolds number $R_{em}^*$ ($R_{em}^* \equiv V_{A0}/v_{dif}^*=V_{A0}/|v_y^*|$ ,hereafter we abbreviate ``effective''). Thus we can treat the magnetic Reynolds number in a context of ideal MHD. 

Here we should note the difference of the definition of the magnetic Reynolds number $R_{em}^* \equiv  V_{A0}/ v_{dif}^*$ (where $ v_{dif}^* \equiv \eta/D$, $\eta$ is the resistivity, $D$ is the actual thickness of the diffusion region) from that of usual one ($ R_{em} \equiv V_{A0}/[\eta/L]$) where $L$ is the system scale. The value of $R_{em}^*$ is considerably smaller than $R_{em}$ ($ R_{em}^* \sim 10^{-7} R_{em} $) since $L \sim 10^7 D$ in typical solar corona.

\subsection{Difference from previous works}
\label{sec:difference}
The main frame of the logic is almost the same with our previous works (papers 3 and 4), but there is one essential difference: Supposing a freedom of the motion of the reconnection point along the current sheet as follows. 

In our previous works, the reconnection point is presumably fixed at the center of symmetry which is in the middle of the current sheet. In this case, very strong converging inflow forms as $R_{em}^*$ increases, and results in a channel flow structure (see figure 7 of paper 4). Finally, the horizontal component of the converging speed exceeds the asymptotic Alfv\'{e}n speed $V_{A0}$ (see figure 4 of paper 4). This is cure because $V_{A0}$ is the maximum speed of the spontaneous flow when all the initial magnetic energy converts to the kinetic energy. The reason of this curious result is probably in the above presumable assumption. 

In order to avoid this difficulty, let us introduce the freedom of the motion of the reconnection point (see section \ref{sec:intro} and subsections 4.1 and 4.6 in paper 4). This improvement enables us to naturally obtain a plausible result: the continuous transition of the structure as a result.

\subsection{Model} 
\label{sec:model}
We present a schematic picture of the reconnection outflow in our self-similar evolution model in figure \ref{fig:structure-sch}. First we introduce two coordinate systems: the fixed coordinate and the zoom-out coordinate. The fixed coordinate is the conventional Cartesian coordinate in which a position vector \mbox{\boldmath $r$} from the origin pointing a point at rest relative to the initial plasma in equilibrium is a constant vector. The zoom-out coordinate is defined as $\mbox{\boldmath $r$}' \equiv \mbox{\boldmath $r$}/[V_{A0} t]$. In the zoom-out coordinate, any self-similarly expanding state looks to be stationary. The coordinate axes of the zoom-out coordinate are defined as follows: $x$-axis is parallel to both the current sheet and initial anti-parallel magnetic field, $y$-axis is perpendicular to the current sheet, and $z$-axis is parallel (the right-handed direction) to the current sheet, but is perpendicular to the initial magnetic field (hence, $\partial_z$=0 in this two-dimensional problem). 

We impose a freedom corresponding to the motion of the reconnection point. The locus of the reconnection point (coincides with the diffusion region) $x=x_y$ is moving along the $x-$axis at the speed $x_y V_{A0}$ relative to the initial plasma in equilibrium because the system is self-similarly expanding. The symmetry with respect to $x-$ and $y-$axes allows us to treat only the region $x,\ y \geq 0$. The outflow is composed of two different plasmas which have different origins. These two plasmas are touch in contact at the point $x=x_c$ (the contact discontinuity). A plasmoid forms around the contact discontinuity. The rear-half region ($x < x_c$: the reconnection jet) is filled with the reconnected plasma coming from outside the current sheet (the inflow region). The front-half region ($x > x_c$: the plasmoid) is filled with the original current sheet plasma. 

From the linear analysis of the inflow region in paper 2, we obtained the result that the inflow velocity is parallel to the $y-$axis in the fixed coordinate (see equations [23] and [24] in paper 2). In order to match the outflow region to the linearized inflow region, we impose a presumable assumption: $x-$component of the inflow velocity vanishes in the fixed coordinate (see figure \ref {fig:fixed-flow-sch}). This directly results in $v_{xp}=-x_y$ at the reconnection point in the zoom-out coordinate. We can clearly understand that, in the zoom-out coordinate, an apparent convergence into the diffusion region occurs owing to the reconnection point motion along the current sheet (see figure \ref {fig:comoving-flow-sch}). We can expect that this apparent converging flow results in a reduction of the reconnection rate as in the case of real converging inflow in our previous work (see section 3 of paper 4). 

When $x_y>0$, the reconnected magnetic flux is stored also in the region $|x|<x_y$. This magnetic flux forms a magnetic island structure between two reconnection points at $x=\pm x_y$. I paid scant attention to this region because we cannot find any clear discontinuity in this region (from our preliminary numerical simulation: not published yet) and hence we cannot formulate as a shock-tube problem. For simplicity, we suppose a uniform distribution of the stored vertical magnetic flux in the island. 

The entire outflow is surrounded by a slow shock which has a complicated `crab-hand' shape (see Abe \& Hoshino 2001). The Petschek-like slow shock (an oblique shock) is elongated from the diffusion region with a slight opening angle $\theta$. When $x_y=0$, the Petschek slow shock has figure-X shape. If $x_y>0$, the reconnection point separates into two points ($x=\pm x_y$). There is a reverse fast shock (an almost perpendicular shock) inside the reconnection jet ($x=x_f$) in some cases. In front of the plasmoid, a forward figure-V-shaped slow shock (an oblique shock) forms. The opening angle and the locus (the crossing point with $x-$axis) are $\phi$ and $x_s$, respectively. 

The entire structure including the several discontinuities is analogous to the one dimensional shock-tube problem. I approximate this reconnection outflow as a quasi one-dimensional problem in order to solve it analytically. Such an approximation may be valid near the $x-$axis, because the system is symmetric with respect to the $x-$axis. Hence we focus our attention on the quasi-one-dimensional problem along the inflow stream line and the reconnection outflow as follows. 

I assume the non-converging inflow seen in the fixed coordinate in order to keep the consistency with the inflow region. This is the critical difference from the previous work (paper 4). I treat the figure L-shaped region $x=x_y$ and $y=0$ (a finite hight up from the current sheet [$finite \gg D_0$ where $D_0$ is the initial current sheet thickness]). Each region between two neighboring discontinuities is approximated to be uniform. 

We note about the up-stream region p just above the Petschek-like slow shock. This region between the slow shock and the separatrix field line (the critical field line reaching the reconnection point: see SFL in figure \ref{fig:comoving-flow-sch}) can be also approximated to be uniform. In this region, each reconnected field line has an almost straight shape and crosses the shock while each field line has hyperbolic shape in the region above the separatrix field line. 

The region 3 between the contact discontinuity and the V-shaped slow shock looks to be non uniform. Since we do not know how we can solve the two dimensional structure of this region analytically, I roughly approximate that the total pressure is uniform even in this region. 

The quantities denoting the initial uniform equilibrium at the asymptotic region are gas pressure $P_0$, mass density $\rho_0$ and magnetic field strength $B_0$. The plasma-$\beta$ value at the asymptotic region is defined as $\beta_0 \equiv P_0/(B_0^2/[2 \mu])$ where $\mu$ is the magnetic permeability of vacuum. In the rest of this paper, we use the normalization of physical quantities as like in paper 2. We define units for each dimension as follows: (unit of velocity)$=V_{A0} \equiv B_0/\sqrt{\mu \rho_0}$ (Alfv\'{e}n speed at the asymptotic region), (unit of the length)$=V_{A0} t$ where $t$ is the time from the onset of reconnection, (unit of mass density)$=\rho_0$, (unit of magnetic field)$=B_0$, (unit of pressure)$=\beta_0/2 \cdot \rho_0 V_{A0}^2$.  

Let us set the quasi-one-dimensional shock-tube-like problem as follows. The system has 23 unknown quantities: $P_p$, $\rho_p$, $v_{xp}$, $v_{yp}$, $B_{xp}$, $B_{yp}$, $\theta$, $P_1$, $\rho_1$, $v_1$, $B_1$, $x_f$, $P_2$, $\rho_2$, $v_2$, $B_2$, $P_3$, $\rho_3$, $v_{y3}$, $B_{x3}$, $B_{y3}$, $\phi$ and $x_s$ where $P_*$, $\rho_*$, $v_*$, $B_*$ denote the pressure, density, velocity, magnetic field, respectively (Note $x_c=v_{x3}=v_2$ because no mass flux passes through the contact discontinuity). The suffices $p$, 1, 2 or 3 denote the region divided by the discontinuities (see figure \ref {fig:structure-sch}). The suffices $x$ or $y$ denote the vector components. Quantities $\theta$, $x_f$, $\phi$ and $x_s$ denote the inclinations of the Petschek-like slow shock, the locus of the fast shock, the inclination of the V-shaped slow shock and the locus of the V-shaped slow shock (crossing point with $x-$axis), respectively. 

These unknowns should be related to each other via conditions coming from the integrated form of conservation laws (i.e., the Rankine-Hugoniot [R-H] conditions) or other relations. The set of relations is listed in the next subsection.

\subsection{Basic Equations} 
\label{sec:b-eqs}
There is no essential difference of the basic equations for the outflow structure between the equations used in paper 4 except that we must include $v_{xp} = -x_y \neq 0$. 

According to the above model of the reconnection outflow, we obtain the following 22 equations for 23 unknown quantities (Detailed forms of each equation are listed in the appendix). I choose $v_{yp}$ as the variable controlled by hand (see section \ref {sec:procedure}). \\

\noindent
1. relations between pre-Petschek-like slow shock and asymptotic region \\
(a) Frozen-in condition (eq. [\ref{eq:frozen-in}])\\
(b) Polytropic relation (eq. [\ref{eq:polytrope}])\\

\noindent
2. R-H jump conditions at Petschek-like slow shock\\
(a) Pressure jump (eq. [\ref{eq:p-p-jump}])\\
(b) Density jump (eq. [\ref{eq:p-d-jump}])\\
(c) Velocity jump (parallel comp.) (eq. [\ref{eq:p-vpara-jump}])\\
(d) Velocity jump (perpendicular comp.) (eq. [\ref{eq:p-vperp-jump}])\\
(e) Magnetic field jump (parallel comp.) (eq. [\ref{eq:p-bpara-jump}])\\
(f) Magnetic field jump (perpendicular comp.) (eq. [\ref{eq:p-bperp-jump}])\\

\noindent
3. R-H jump conditions at reverse fast shock\\
(a) Pressure jump (eq. [\ref{eq:f-p-jump}])\\
(b) Density jump (eq. [\ref{eq:f-d-jump}])\\
(c) Velocity jump (eq. [\ref{eq:f-v-jump}])\\
(d) Magnetic field jump (eq. [\ref{eq:f-b-jump}])\\

\noindent
4. Magnetic flux conservation at reconnection point (eq. [\ref{eq:bflux-cons}])\\

\noindent
5. Force balance at contact discontinuity (eq. [\ref{eq:f-balance}])\\

\noindent
6. R-H jump conditions at forward V-slow shock\\
(a) Pressure jump (eq. [\ref{eq:s-p-jump}])\\
(b) Density jump (eq. [\ref{eq:s-d-jump}])\\
(c) Velocity jump (parallel comp.) (eq. [\ref{eq:s-vpara-jump}])\\
(d) Velocity jump (perpendicular comp.) (eq. [\ref{eq:s-vperp-jump}])\\
(e) Magnetic field jump (parallel comp.) (eq. [\ref{eq:s-bpara-jump}])\\
(f) Magnetic field jump (perpendicular comp.) (eq. [\ref{eq:s-bperp-jump}])\\

\noindent
7. Boundary Condition at the tip of the outflow (eq. [\ref{eq:bc-tip}])\\

\noindent
8. Magnetic flux conservation all over the outflow (eq. [\ref{eq:bflux-cons2}])\\

We can solve the most of these equations by hand, and by substituting the solutions into other equations, we can reduce equations. Finally, ten equations [\ref{eq:frozen-in}], [\ref{eq:p-p-jump}], [\ref{eq:p-d-jump}], [\ref{eq:p-vpara-jump}], [\ref{eq:p-vperp-jump}], [\ref{eq:p-bpara-jump}], [\ref{eq:f-v-jump}], [\ref{eq:f-balance}], [\ref{eq:bc-tip}] and [\ref{eq:bflux-cons2}] remain as complicated nonlinear coupled equations for ten unknowns $\rho_p$, $v_{xp}$, $B_{xp}$, $B_{yp}$, $\theta$, $P_1$, $\rho_1$, $x_f$, $\phi$ and $x_s$ with a controllable variable $v_{yp}$.

\subsection{Numerical procedure}
\label{sec:procedure}
We solve these coupled ten equations by an iterative method (the Newton-Raphson method). In order to find the well converged solution of the highly nonlinear coupled equations, a precise initial guess of the unknowns is required. In general, the step to find an appropriate initial guess is the core of difficulty (this should be comparable to a ``treasure hunting in ten-dimensional space'' with an incomplete treasure map). Fortunately, this most difficult step had already been cleared in paper 4 for the case $x_y=0$. 

We can start from the solution of the case $x_y=0$ for an arbitrary value of $\beta_0$. Our most interest is focused on low-$\beta$ cases typically encountered in astrophysical problems. We demonstrate the numerical procedure for the case $\beta_0=0.01$ as the reference case because this is a typical value in the solar corona and geomagnetosphere. In paper 4, we have already had the solution in the spontaneous maximum inhalation case ($x_y=0$) for $\beta_0=0.01$.

First, we start with the solution of the case $v_{yp} = -v_{insp}(\beta_0=0.01) = -0.06398$ (where $v_{insp}[\beta_0=0.01]$ is the spontaneous maximum inhalation speed for $\beta_0=0.01$) as the initial guess for the case $v_{yp}=-v_{insp}(0.01)-\Delta v_{y}$, where $\Delta v_{y}$ is the decrement of $v_{yp}$ (see paper 3). Once we find a converged solution of the Newton-Raphson procedure for the case $v_{yp}=-v_{insp}(0.01)-\Delta v_{y}$, we treat it as the initial guess for the case $v_{yp}=-v_{insp}(0.01)-2 \Delta v_{y}$, and we have successively obtained the series of solutions for different $v_{yp}$. Of course, $\Delta v_{y}$ should be a small enough value to keep good convergence of the Newton-Raphson method. From several trials, I carefully adopt the decrement: $\Delta v_{y}=10^{-8}$. 

We obtain the result that, as $|v_{yp}|$ increases, the locus  $x_y$ of the reconnection point drastically increases (see figure \ref{fig:xy}: this means that the reconnection point is moving in $x-$direction at a speed $x_y V_{A0}$). Simultaneously the strength of the reverse fast shock decreases. At a critical value $|v_{yp}|= -0.063975$, the reverse fast shock no longer forms (the density jump ratio and the pressure jump ratio reduce to unity). At this point we exchange the numerical code to another one for the case with no reverse fast shock forms (see the last part of appendix), and we can continue the calculation. Above procedure is valid for any value of $\beta_0$.

The above scan of $v_{yp}$ continues as long as we can obtain effectively converged solutions. I performed the calculation with 16 digit degree of complex numbers. I set the convergence goal of the Newton-Raphson method to be 10 digit degree of precision. In this case, $R_{em}^* \sim 2100$ (where $R_{em}^* \equiv B_{xp}/(v_{xp} B_{yp}-v_{yp} B_{xp})$: the normalized \mbox{\boldmath $E$}$\times$\mbox{\boldmath $B$}-drift velocity) is almost the upper limit of the magnetic Reynolds number for efficient successive calculation. If we try further calculation, the Newton-Raphson method requires much smaller $\Delta v_{y}$ and much more digit degree of numbers. It is so time consuming that we cannot efficiently proceed to much higher $R_{em}^*$ regime. Fortunately, since we have already obtained sufficiently settled asymptotic behavior as shown in the following,  I believe the further calculation is not worthy.

\subsection {Summary of procedure}
\label{sec:sum-pro}
We here summarize our procedure. We solve the outflow region approximated as a quasi-one dimensional shock tube problem along the figure-L shaped line ($x=x_y$ and $y=0$). Each region between neighboring discontinuities is approximated to be uniform for simplicity. The system is described by two parameters: the plasma-$\beta$ $\beta_0$ in the asymptotic region and the magnetic Reynolds number $R_{em}^*$. We fix $\beta_0=0.01$ as a typical value for the solar corona. This problem has 23 unknown quantities: $P_p$, $\rho_p$, $v_{xp}$, $v_{yp}$, $B_{xp}$, $B_{yp}$, $\theta$, $P_1$, $\rho_1$, $v_1$, $B_1$, $x_f$, $P_2$, $\rho_2$, $v_2$, $B_2$, $P_3$, $\rho_3$, $v_{y3}$, $B_{x3}$, $B_{y3}$, $\phi$ and $x_s$. We treat $v_{yp}$ as a controllable variable handled by our hand. We can obtain other 22 quantities by solving 22 coupled equations (\ref {eq:frozen-in})-(\ref{eq:bflux-cons2}) denoting junction conditions over discontinuities. 

The magnetic Reynolds number (the key parameter of this paper) is estimated by normalized \mbox{\boldmath $E$}$\times$\mbox{\boldmath $B$}-drift velocity as $R_{em}^* \equiv B_{xp}/(v_{xp} B_{yp}-v_{yp} B_{xp})$ by using obtained solution for unknowns. In the followings, the results are shown as functions of $ R_{em}^*$.

\section{Result}
\label{sec:res}
\subsection{Reduction of the reconnection rate}
\label{sec:res-rec-rate}
Our major interest is focused on the reconnection rate. Figure \ref{fig:rec-rate} shows the magnetic Reynolds number ($R_{em}^*$) dependence of the reconnection rate ($R^*$) derived from the above approximated shock-tube problem in a very wide dynamic range of $R_{em}^*$ ($15 < R_{em}^* < 2100$) where $R^*$ is defined as the normalized reconnection electric field $R^* \equiv v_{xp} B_{yp}-v_{yp} B_{xp}$ in the unit of $V_{A0} B_0$. We can clearly find the asymptotic dependence $R^* \propto (R_{em}^*)^{-1}$.

We should note again here that the definition of the magnetic Reynolds number $R_{em}^*$ adopted in this work is different from usual one: the quantity having dimension of the length included in the definition is not the system scale but the thickness of the diffusion region (see section \ref{sec:concept}). Hence the value of $R_{em}^*$ is almost seven orders of magnitude smaller than usual one for the same resistivity if we estimate the thickness of the diffusion region to be of the order of the ion Larmor radius ($\sim 10^0$ [m] in typical solar corona). 

This reduction of the reconnection rate is not owing to the reduction of the inflow speed $v_{yp}$, but to the enhancement of the apparent convergence of the inflow toward the reconnection point arising from the motion of the reconnection point (see figure \ref{fig:quantities}). Even though the vertical inflow speed $|v_{yp}|$ is almost constant, the apparent horizontal inflow speed (equals to the moving speed of the reconnection point) $|v_{xp}|$ increases as $R_{em}^*$ increases. The direction of the inflow velocity tends to be parallel to that of the magnetic field lines as $R_{em}^*$ increases (see figure \ref{fig:tilt}). This results in a reduction of the reconnection electric field, thus the reconnection rate reduces (remember that the reconnection rate is non-dimensional normalized value of the reconnection electric field).

The reduction mechanism of the reconnection rate is almost the same with that of my previous work (see paper 4), but there is one crucially different point. In paper 4, the actual converging speed $|v_{xp}|$ exceeds the ambient Alfv\'{e}n speed $V_{A0}$ when $R_{em}^*$ sufficiently increases. This is unrealistic because the spontaneous inflow speed cannot exceed the Alfv\'{e}n speed. The apparent converging speed of this work asymptotically settles to the ambient Alfv\'{e}n speed for very large $R_{em}^*$. We must note that the direction of the magnetic field and that of the inflow velocity tends to be parallel even if $|v_{xp}|$ is bound for the ambient Alfv\'{e}n speed (see  figure \ref{fig:tilt}). 

We find that the motion of the reconnection point results in the reduction of the reconnection rate. This means that the variation of the reconnection rate is owing to the transition of the structure of the reconnection system. This is the most important property of this model with moving reconnection point. All of these changes are caused by the variation of the magnetic Reynolds number $R_{em}^*$.

\subsection{Continuous transition of the inflow structure} 
\label{sec:res-transition}
This work clarifies the continuous transition of the entire structure from the Petschek-like one having large reconnection rate to the double reconnection point one having small reconnection rate in a series of solutions. When the magnetic Reynolds number is small enough ($R_{em}^* \sim 15.63$ for $\beta_0=0.01$: hereafter our discussion is limited to the representative case $\beta_0=0.01$ because this is a typical value in the solar corona), a single reconnection point forms at the origin of the coordinate. When $R_{em}^* > 15.63$, the locus $x_y$ of the reconnection point shifts to the both directions along the outflow and the system forms double reconnection points at $ x=\pm x_y$. Also a magnetic island forms between them (see figure \ref{fig:xy}). Note that we do not solve the region $|x|<x_y$ by the shock-tube analysis. Instead of it, we assume a uniform distribution of the reconnected magnetic flux in this region for simplicity (see section \ref{sec:model} ). 

The shock-tube analysis provides the boundary condition to solve the inflow region by the Grad-Shafranov (G-S) method (see sections 3.2 and 3.3 of paper 2). We impose this boundary condition on $y=0$ for simplicity. Of course, strictly to saying, this boundary condition should be imposed along the edge of the reconnection outflow, not on $y=0$. The reasons to accept this simplicity are the followings: 1) The shape of the edge is rather complicated to be treated by simple way, 2) Because the boundary condition obtained here is based on a drastic simplification (the quasi-one diminsional shock-tube approximation), it should be meaningless even if only here pursues strictness. 

By solving the G-S equation by a SOR routine (see paper 2), we can obtain the inflow structure corresponding to individual outflow solution obtained above. The resultant inflow structure continuously changes according to the variation of $R_{em}^*$: A typical Petschek-like structure ($R_{em}^* \sim 15.63$: see figure \ref{fig:single-X}), a X-O-X type (figure \ref{fig:X-O-X}) or a double-Y point type solution (figure \ref{fig:double-Y}) for $R_{em}^* > 15.63$. We must note that only the region $x^2+y^2<1$ is solved by the above G-S method in figures \ref{fig:single-X}-\ref{fig:double-Y}. The region $x^2+y^2>1$ keeps its initial condition (an anti-parallel uniform equilibrium configuration) and is not influenced by the reconnection yet. 

Whether the region $|x|<x_y$ is occupied by an island or a collapsed current sheet-like structure is roughly determined by the ratio of the reconnected magnetic field strength ([reconnected magnetic flux]/ $x_y$) to the asymptotic magnetic field strength ($B_0$). If the ratio is greater than unity, an island forms, else (the ratio $\ll 1$) a collapsed current sheet-like structure forms. The critical magnetic Reynolds number is roughly estimated as $R_{em}^* \sim 16.9$ from the shock-tube calculation.

\subsection{Reducible fraction of the initial magnetic energy}
\label{sec:res-reducible}
The solutions for $R_{em}^* > 15.63$ include a magnetic island or a current sheet-like structure (extremely elongated and collapsed island) having a finite amount of magnetic energy. This means that the reconnection can release only a finite fraction of the total magnetic energy of the initial current sheet system by one evolution of the self-similar reconnection. 

Let us estimate the reducible fraction of the initial magnetic energy by single reconnection. Consider the separatrix stream line (SSL in figure \ref{fig:reducible-sch}) reaching the reconnection point in the zoom-out coordinate (including the apparent converging flow). We approximate the shape of the separatrix stream line to be straight for simple estimation. All the magnetic flux in the region below the separatrix stream line can reach the Petschek-like slow shock and its magnetic energy will be released. We approximate here that the magnetic energy of the plasma which enters the Petschek-like slow shock is completely converted to the kinetic / thermal energy of the reconnection outflow for simple estimation. By noting that this structure is self-similarly expanding, we can calculate the reducible fraction of the initial magnetic energy as follows. 

The system is initially just like a point at the origin in the conventional fixed coordinate. The system starts to expand self-similarly and sweeps the plasma in the initial current sheet system. At an instant, the region below a straight line (dashed line in figure \ref{fig:reducible-sch}) with inclination angle $\theta_R$ from the initial current sheet is the region already swept by the expanding past separatrix SSL or the region below the present SSL. Hence the magnetic energy in the triangle-shaped region ($0 \leq \theta \leq \theta_R$ in figure \ref {fig:reducible-sch}) is reducible. We call this the ``reducible region''. The angle $\theta_R$ is obtained as
\begin{equation}
\theta_R \equiv \tan^{-1} [(1-x_y)(v_{yp}/v_{xp})]\ .
\end{equation}
Thus the reducible fraction $f$ is obtained as 
\begin{equation}
f=\theta_R/(\pi/2)\ .
\end{equation}
The magnetic Reynolds number dependence of $f$ is shown in figure \ref{fig:reducible}. We can see that the reducible fraction drastically decreases as the magnetic Reynolds number increases (asymptotically $f \propto {R_{em}^*}^{-1/2}$). 

The intuitive meaning of the reducible fraction $f$ is as follows. Let us consider the flow structure in the view point of the zoom-out coordinate. Note that the apparent converging inflow appears even in the region outside the FRWF in which there is no real flow in the fixed coordinate (see figure \ref {fig:reducible-sch}). All the stream lines in the region outside the FRWF radially converge toward the origin. We should note that only the plasma at the lower latitude region ($\leq \theta_R$: the reducible region) from the $x-$axis (the original current sheet) reaches the region below the SSL when this plasma is swallowed to the expanding FRWF. This plasma can enter the  Petschek-like slow shock within a finite time. On the contrary, the plasma at the higher latitude region ($> \theta_R$) does not enter the Petschek-like slow shock and accumulates into the magnetic island or the new current sheet (see figures \ref{fig:X-O-X}  and \ref{fig:double-Y} ). Thus the magnetic energy only in the reducible region ($\leq \theta_R$) can be released by the self-similarly expanding reconnection system.

\section{Summary and discussion}
\label{sec:sum-dis}
\subsection{Summary}
\label{sec:sum}
We had investigated a magnetic reconnection in a free-space which is free from any external influences in a series of papers (see papers 1-4). This paper is an extension to more universal situation. The new point of this study is introducing a freedom of the motion of the reconnection point along the initial current sheet. I found the following major results. 

1) Reconnection rate\\
The reconnection rate $R^*$ is a decreasing function of the magnetic Reynolds number $R_{em}^*$. This tendency is a natural result of the motion of the reconnection point. At the lowest value of the magnetic Reynolds number ($R_{em}^*=15.63$ for the case $\beta_0=0.01$), the single reconnection point forms at the origin. As $R_{em}^*$ increases, the locus of the reconnection point ($|x|=x_y$ in the zoom-out coordinate) increases (see figure \ref{fig:xy}). This means that the reconnection point is moving along the original current sheet at a speed $x_y V_{A0}$. This motion of the reconnection point naturally leads an apparent convergence of the inflow toward the reconnection region in the zoom-out (i.e., the reconnection point-comoving) coordinate (see figure \ref {fig:comoving-flow-sch}). Thus the reconnection rate (the normalized electric field) reduces because the directions of the inflow velocity and the magnetic field tend to be parallel in the zoom-out frame (see figure \ref{fig:tilt}).

2) Continuous transition of solutions\\
When $x_y=0$, the solution is very similar to the original Petschek model and having the largest reconnection rate (typical fast reconnection). When $x_y>0$, two reconnection points form at $x=\pm x_y$. Between these reconnection points, a magnetic island or a new current sheet (exactly to saying, it is an extremely elongated and collapsed magnetic island) forms. Hence, the system has a typical X-O-X structure or a double Y-point structure. This transition of the structure has close relation to the reduction of the reconnection rate as discussed above. As $R_{em}^*$ increases, $x_y$ also increases. The structure continuously changes to the X-O-X type or the double Y-point type having smaller reconnection rate (slower reconnection). We can conclude that the structure directly corresponds to the speed of the energy conversion. These structures expand self-similarly until some external circumstance influences the spontaneous evolution of the reconnection system. 

3)Reducible fraction of magnetic energy\\
The reducible fraction $f$ of the initial magnetic energy of the current sheet system is restricted in relation to the structures of the reconnection system in the above discussion. As $x_y$ increases (as $R_{em}^*$ increases), $f$ decreases (see figure \ref{fig:reducible}). When the system has the X-O-X structure or the double Y-point structure, these structures remain the magnetic energy in the island or the newly formed current sheet-like structure. This means that a reconnection having these structure can release only a finite fraction of the magnetic energy of the initial current sheet by one reconnection even though the current sheet magnetic energy is fully a free energy. 

In the following part, let us discuss the physical meanings and the applications of the above results.

\subsection{Reconnection rate}
\label{sec:dis-rec-rate}
We obtained the magnetic Reynolds number dependence of the reconnection rate as $R^* \propto (R_{em}^*)^{-1}$. This behavior of the reconnection rate is closely associated with the motion of the reconnection point. When $x_y>0$, the reconnection point is moving at the speed $x_y V_{A0}$ in the direction parallel to the initial current sheet. In this work, the inflow direction is supposed to be perpendicular to the initial current sheet in the fixed coordinate to keep consistency with the result of our linear analysis in the inflow region (see paper 2). In the zoom-out (reconnection point-comoving) coordinate, the inflow is oblique owing to the apparent horizontal component of the inflow velocity (see figure \ref {fig:comoving-flow-sch}). This apparent motion works as the converging inflow in paper 4 (see section 3 and figure 5). Hence we call this the apparent converging inflow. 

Figure \ref{fig:quantities} shows that absolute value of the horizontal component of the apparent converging inflow does not exceed the ambient Alfv\'{e}n speed $V_{A0}$ (normalized as unity). This is a drastic change from the previous work (see figure 4 of paper 4). Needless to say, the ambient Alfv\'{e}n speed is the maximum value when all the magnetic energy is converted to the plasma bulk kinetic energy. Hence $V_{A0}$ is the maximum speed of any spontaneous flow. The current result ($|v_{xp}| \leq 1$ even if $R_{em}^* \gg 1$) is plausible to be a realistic solution. 

Note that the curve of figure \ref{fig:rec-rate} terminates at $R_{em}^*=15.63$ for the case $\beta_0=0.01$. The author has the following somewhat speculative intuition about the dynamic behavior of the diffusion region that is based on our experience of a lot of numerical simulations in paper 1.  If we impose a very large electric resistivity in order to realize a situation with a small magnetic Reynolds number ( $R_{em}^*<15.63$ ), the thickness of the diffusion region increases as time proceeds, thus the actual diffusion speed cannot increase almost at all. On the contrary, if we impose a very small resistivity to realize a situation with a large magnetic Reynolds number, the thickness of the diffusion region decreases as time proceeds (if the mesh size is sufficiently small). The actual diffusion speed cannot decrease almost at all. This may show a self-regulation mechanism of the diffusion speed. The resultant reconnection rate is very insensitive to the variation of the imposed resistivity. Such dynamical response of the diffusion region had been empirically confirmed in our numerical MHD simulations. 

This dynamic behavior is mainly determined by the spontaneous inhalation speed by the Petschek-like slow shock (see section 5.4 of paper 3). The maximum spontaneous inhalation speed is presented in figure 2 of paper 4. The value for $\beta_0=0.01$ is $|v_{yp}| = 0.06398$ corresponding to $R_{em}^* = 1/ |v_{yp}| =15.63$ (roughly to saying, this corresponds to $10^6$ times the Spitzer resistivity or larger). This means that the magnetic diffusion speed cannot exceed $ 0.06398 V_{A0}$. Even if the resistivity is very large, the diffusion speed will be self-regulated to this maximum value by spontaneously adjusting the thickness of the diffusion region. This clearly shows a passive nature of the diffusion region (see 4.2 of paper 4). 

This speculative discussion is consistent with the logarithmic (extremely weak) dependence of the reconnection rate on the magnetic Reynolds number in the original Petschek model. Such passive behavior of the diffusion region is actual for the case of sufficiently large resistivity in which the diffusion region thickness is much larger than microscopic proper scales, e.g., the ion Larmor radius. In this case, the actual diffusion speed is almost constant independent of the resistivity, hence the magnetic Reynolds number $R_{em}^*\ (=15.63)$ (note that the definition of $R_{em}^*$ is different from usual one; see section \ref{sec:concept} ) and the reconnection rate ($R^*=0.05$) do not vary. This corresponds to the terminal point in the upper left end of the curve in figure \ref{fig:rec-rate}. However, when the resistivity is very small and the diffusion thickness is comparable to its minimum value (e.g., the ion Larmor radius), the response of the reconnection system drastically changes to another scheme discussed in section \ref{sec:res-rec-rate}.

On the contrary, if we force to enhance the inflow speed $v_{yp}$, we obtain the curious results with $x_y<0$ for $|v_{yp}| > 0.06398$. This is nonsense because it does not satisfy the realistic condition $x_y \geq 0$. This may mean the possibility of the diverging inflow, but this is also nonsense because the spontaneous inhalation cannot make any diverging inflow by the fast-mode rarefaction (see Vasyliunas 1975). Thus I believe that we must terminate the curve in figure \ref{fig:rec-rate} at $R_{em}^*\ (=15.63)$ as like in this figure.

We should note that such behavior in very large magnetic Reynolds number cases are very hard to be studied by MHD numerical simulations with finite sized mesh method. If we try to simulate this case, the diffusion region thickness thins to be almost equivalent to the mesh size. This is obviously nonsense because unrealistic numerical diffusion dominates the physical diffusion. This shows the limit of application of the MHD numerical simulation. 

The author is partially skeptical to the original Petschek's discussion on the logarithmic dependence of the reconnection rate to the magnetic Reynolds number in extremely high magnetic Reynolds number region. Petschek might presumably assume that the diffusion region thickness can be infinitesimally thin in extremely high magnetic Reynolds number region to keep self-regulation process of the diffusion speed. We must note, however, that in the actual current sheet system, the diffusion region thickness is bound at the smallest value corresponding to the microscopic proper scale (e.g., the ion Larmor radius). After the diffusion region thickness reaches its minimum value (of the order of the ion Larmor radius), the reconnection rate may not keep its value (as like in the very weak logarithmic dependence) but may decrease more rapidly (may be a power law) as like in the result of this work.

\subsection{Continuous transition from fast regime to slow regime} 
\label{sec:dis-transition}

We have obtained a wide variety of the solutions which have different values of $x_y$ corresponding to the variation of $R_{em}^*$. The transition of the structure results in the variation of the reconnection rate (see figure \ref{fig:rec-rate}). The most significant new point of this work is is to clarify the continuous transition from the fast regime to the slow regime. If the resistivity is large enough ($R_{em}^* = 15.63$), the Petschek-like fast regime takes place (reconnection rate $R^* \sim 0.05$). As the resistivity decreases, the structure changes to the X-O-X type and finally reaches the double Y-point type ($R^* \sim 10^{-7}$). 

Another work which argue the continuous transition from fast regime to slow regime is Priest \& Forbes (1986). In their case, the reconnection solutions compose a family which is parametrized by constants denoting the external boundary condition at the inflow boundary (see section 3 of their paper). Hence their case is clearly based on the so-called ``driven'' model. On the other hand, in our case, the solution changes continuously according to the magnitude of the electric resistivity which may relate to intrinsic properties of the plasma in the system. This work first clarifies the continuous transition based on the spontaneous model. 

Although the inverse scaling of the reconnection rate $R^*$ with $R_{em}^*$ is of the type normally associated with super-slow reconnection (Priest \& Forbes 2000), one must keep in mind that $R_{em}^*$ is an effective magnetic Reynolds number based on the thickness of the diffusion region and not the global scale of the configuration. The reconnection here is truly fast for small $R_{em}^*$ because the electric field at the reconnection point is $\sim 0.05$ of Alfv\'{e}n electric field defined by $V_{A0} B_0$, and is many orders of magnitude greater than would be the case for super-slow or even slow reconnection. If we extrapolate our result to extremely large magnetic Reynolds number cases, $R^*$ decreases to $\sim 10^{-7}$ for the value of $R_{em}^*$ corresponding to the Spitzer resistivity. This is just a typical value for the slow regime, i.e., the Sweet-Parker model. 

This drastic decrement of the reconnection rate is closely related to the transition of the structure of reconnection system. Figures \ref{fig:single-X}-\ref{fig:double-Y} clearly show the transition from the Petschek-like structure (single X-point) to the double-Y structure via the X-O-X structure. During this transition, the locus of the reconnection points ($x=\pm x_y$) separates in both direction along the initial current sheet at a finite speed ($x_y V_{A0}$). The expected structure for extremely large magnetic Reynolds number is the double Y-point structure in which 1) two reconnection points are moving in counter directions at the asymptotic Alfv\'{e}n speed $V_{A0}$, and 2) a new current sheet-like structure forms between two reconnection points. Thus we can find a new solution for the slow reconnection regime which is morphologically very similar to the Sweet-Parker model especially in the inflow region. In addition, its reconnection rate ($\sim 10^{-7}$) is just similar to that of the Sweet-Parker model. We cannot distinguish them not only by the morphological structures but also by the energetics. 

However we should notice the critical differences between them. In our model, even in the case of small reconnection rate, 1) the diffusion region is localized around the Y-point (c.f., spread out into the entire current sheet in the Sweet-Parker model), 2) the magnetic energy is converted by the Petschek-like slow shock which elongates from the Y-point (c.f., converted by the magnetic diffusion in the entire current sheet in the Sweet-Parker model). When thinking from the view point of the reconnection rate, this is typical slow reconnection like the Sweet-Parker solution. However, when thinking from the view point of the energy conversion mechanism, it is considered as an extreme limit of the smooth transition from the fast regime as the Petschek solution, because the magnetic energy is mainly converted by the slow-mode wave (not the magnetic diffusion). Therefore, we must consider this is a quite new type of slow reconnection solution.

\subsection{Reducible fraction of magnetic energy by one reconnection} 
\label{sec:dis-reducible}
The majority may think that the magnetic energy stored in the current sheet system can be fully released by a reconnection. However, in section \ref{sec:res-reducible}, we have clearly shown the magnetic Reynolds number dependence of the reducible fraction of the initial magnetic energy. If the electric resistivity decreases, released fraction of the magnetic energy by single reconnection also decreases (see figure \ref{fig:reducible}). We can find that the magnetic energy of the current sheet system is analogous to the free energy of thermodynamic systems: the reducible fraction is restricted by additional conditions (e.g., the resistivity in this case). 

Let us consider an ensemble of equivalent current sheet systems. While each configuration and stored magnetic energy is the same, the released energy can differ depending on the individual value of the resistivity enhanced in the diffusion region. If the resistivity is large enough and the Petschek-like reconnection takes place, the magnetic energy will be completely released. On the contrary, if the resistivity is small and the double Y-point reconnection takes place, released energy will be very small. Finally the reducible fraction $f$ tends to $f \propto {R_{em}^*}^{-1/2}$. 

We should note that it is insufficient to estimate the magnitude of released energy by single reconnection even if we can estimate the detailed configuration and the magnetic strength of the current sheet system. The released energy strongly depends on the magnetic Reynolds number which is very difficult to forecast beforehand.

\subsection{Observational estimation of reconnection rate}
\label{sec:rec-rate-obs}
In a lot of works, the reconnection rate is defined as the inflow Alfv\'{e}n Mach number. This definition, however, is not appropriate for the case having converging inflow as like in this study. We must note that the inflow Alfv\'{e}n Mach number itself does not denote the reconnection rate if the inflow is inclined with respect to the magnetic field lines in the zoom-out (reconnection point-comoving) frame. As an extreme example, if the inflow velocity is just parallel to the magnetic field lines, the reconnection rate is vanished even if the inflow speed is not vanished. Since the \mbox{\boldmath $E$}$\times$\mbox{\boldmath $B$}-drift determines the shift speed of the magnetic field line, the actual reconnection rate should be defined by the reconnection electric field not by the inflow speed: $R^* \equiv v_{xp} B_{yp}- v_{yp} B_{xp}$. The popular estimation of the inflow  Alfv\'{e}n Mach number for the reconnection rate coincides with true reconnection rate only if $v_{xp}=0$ (no converging flow or fixed reconnection point) and $ B_{xp}=1$ (same with ambient magnetic field strength). 

Recent observation enables us to estimate the reconnection rate for actual reconnection events. Tsuneta (1996) first estimated the reconnection rate (the inflow Alfv\'{e}n Mach number $\sim 0.07$) in a solar flare event on 1992 Feb. 21 by Yohkoh observation. Yokoyama et al. (2001) estimated the inflow Alfv\'{e}n Mach number ($\sim 0.001-0.03$) in a flare on 1999 March 18 by SOHO observation. Isobe et al. (2005) and Nagashima \& Yokoyama (2006) also estimated  the inflow Alfv\'{e}n Mach number for several flares. If we note significant ambiguity to estimate the Alfv\'{e}n speed, we may conclude that these estimated values of the reconnection rate are well consistent with almost theoretical upper-limit value (of the order of $10^{-2}$) of the fast reconnection models including the self-similar evolutionary model. 

However we should note that these studies can make sense to estimate the reconnection rate only if the inflow is approximately perpendicular to the magnetic field lines (realistically saying, it is plausible only if the reconnection point is at rest in the initial plasma-rest frame). If the reconnection point is moving at a considerable speed comparing with the Alfv\'{e}n speed, we must adopt the correct definition $R^* \equiv v_{xp} B_{yp}- v_{yp} B_{xp}$ for estimation instead of the inflow Alfv\'{e}n Mach number. For example, this is important for the case of the loop-top flare events in which the reconnection point will rise upward at a considerable speed.

\subsection {Relation to particle acceleration}
The reconnection system may play an important role also for the particle acceleration. First, Tsuneta and Naito (1998) pointed out that the reconnection system having reverse fast shock in the outflow surrounded by the Petschek-like slow shock can work as an efficient electron accelerator. In their result, non-thermal electrons (20-100 KeV) can be produced in a very short time (0.3-0.6 sec) by the first order Fermi acceleration process around a quasi perpendicular fast reverse shock. A theoretical extension by Mann et al. (see Mann et al. 2006) argues the electron acceleration in the same situation by the shock drift acceleration up to 10 MeV. 

We must note that, in the self-similar evolutionary reconnection model, the parameter range in which the reverse fast shock forms in the reconnection outflow is not so wide (low $\beta$: $\beta < 0.441$ [see figure 2 of paper 4], not so high magnetic Reynolds number: $R_{em}^* < 22.9$ [see figure \ref{fig:rec-rate}]). If the parameters are in this range, the self-similar reconnection system also can work as an efficient particle accelerator. Because the acceleration timescale ($\sim 10^{-1}$[sec]) estimated by Tsuneta \& Naito (1998) is much shorter than the evolution timescale (the Alfv\'{e}n transit time $\sim 10^2$[sec] for typical solar flares) of the reconnection system, the energy spectrum of the accelerated particles may be conserved during the evolution while the total amount (particle flux) of the accelerated particles increases as the system scale increases in proportion to the time from the onset.

\subsection{Relation to micro/nano flares}
\label{sec:dis-nano-flare}
The property discussed in section \ref{sec:dis-reducible} for the cases of very large magnetic Reynolds number ($R_{em}^* \gg 1$) inspires us a relation to micro- or nano-flares. As a result in section \ref{sec:res-reducible}, the reconnection with very small resistivity can release only a small fraction of the magnetic energy of the initial current sheet system at one evolution even if the stored magnetic energy is huge. This may correspond to small energy-scale flares. 

As discussed in section \ref{sec:dis-transition}, the release of the magnetic energy takes place on the slow shock elongated from the reconnection point. The energetic region (here after, the ``hot region'') around $x \sim \pm x_y$ is moving along the initial current sheet at the speed $\sim x_y V_{A0}$ where the locus $x_y\ (<1)$ of the reconnection point is an increasing function of $R_{em}^*$. These hot regions will be observed as a couple of two bright points (in X-ray or UV depending on the temperature) having a bipolar motion along the magnetic field. It will be seen as phenomena that bright points spread along the magnetic flux tube (typically seen in the so-called ``loop flares'', e.g., in the TRACE movies of C6.3 flare on 2000 Sep. 5 or C6.1 flare on 2001 Aug. 26).

If the magnetic Reynolds number is somewhat larger ($R_{em}^* \sim 15.6-20$; see section \ref{sec:res-transition}) than that of the single X-point case, the remarkable X-O-X type reconnection occurs (see figure \ref{fig:X-O-X}). The magnetic energy remained in the magnetic island cannot be released by further magnetic reconnections because the remained island will be stable for magnetic reconnections. Hence this case is one-off phenomena which can release a finite fraction of the magnetic energy ($\geq$ 36\% of the initial magnetic energy of the current sheet; from figure \ref{fig:reducible}). The restriction of the released fraction of the magnetic energy is related to the drastic topology change of the magnetic field lines (i.e., from ``current sheet'' to ``island''). 

The very large magnetic Reynolds number cases  ($R_{em}^*  \gg 20$) which form the double Y-point type structure are more interesting. An extremely elongated magnetic island remains between the two Y-points. This collapsed island looks like a new current sheet which will be unstable again for reconnections; see figure \ref{fig:double-Y}. The author emphasize that this double Y-point type reconnection with moving hot regions can recurrently occur in a current sheet system. Even if the amount of released energy by one reconnection is extremely small, such recurrent occurrence of the double Y-point type reconnections finally can release a large fraction of the magnetic energy of the initial current sheet system. In addition, this type of reconnections may occur simultaneously to form a nested spatial self-similar structure. 

The first-light image of HINODE (Solar-B satellite) revealed a lot of very small X-ray bright points simultaneously take place on the entire solar surface. This phenomenon might correspond to the reconnections at very high magnetic Reynolds number. Such small, short time and faint events which we were not able to detect before HINODE might contribute to the coronal heating. In that case, we might be able to obtain the answer to the long term open question ``coronal heating problem'' soon.

\vspace{1cm}

The author thanks NAOJ Solar physics group (Saku Tsuneta, Kuniko Hori, Shinsuke Imada and Suguru Kamio), Kazunari Shibata (Kyoto Univ.), Takaaki Yokoyama, Masahiro Hoshino (Tokyo Univ.), NAOJ Theoretical group (Takahiro Kudoh and Mami Machida), Ritoku Horiuchi (NIFS) and Kanya Kusano (Earth Simulator Center of JAMSTEC) for fruitful scientific discussion and comments. Atsuhiro Nishida (SOKENDAI)'s comment on the discussion of sections \ref {sec:res-reducible} and \ref{sec:dis-reducible} was very effective to revise. Improvement of English usage is owing to Naoko Kato (SOKENDAI). The numerical calculation and the edition of the paper were performed by notebook PCs that borrowed from SOKENDAI. 

Finally the author would like to show my special thank to Takeo Kosugi (ISAS/JAXA) who was the project manager of HINODE (Solar-B) team for his successive encouragement throughout the series of my works. He did not wait the scientific result of the mission and passed away just after the success of the first-light of HINODE.

\appendix
\section{Equations for structure of the reconnection outflow}

The structure of the reconnection outflow is determined by the following equations. \\

We assume that the region between the asymptotic region and the pre-shock region is filled with non-resistive plasma. Hence, the magnetic flux is frozen into the induced inflow. We also assume a polytropic variation in the induced inflow because there is no violent process in the fast-mode rarefaction. Each region between neighboring discontinuities is approximated to be uniform. 

The essential difference from our previous works (papers 3 and 4) is introducing a new unknown $x_y$ (the locus of the reconnection point). However, we do not need new physical equation by assuming $x_y=-v_{xp}$ (see section \ref{sec:b-eqs}) in order to keep consistency with the linear perturbation theory in the inflow region. The alterations are in equations [\ref {eq:p-vpara-jump}], [\ref {eq:p-vperp-jump}], [\ref {eq:bflux-cons}], [\ref {eq:bc-tip}] and [\ref {eq:bflux-cons2}] as followings. 

Thus, we impose the frozen-in condition 
\begin{equation}
\frac{B_0}{\rho_0}=\frac{B_{xp}}{\rho_p} \ ,\label{eq:frozen-in}
\end{equation}

and the polytropic relation 
\begin{equation}
P_0 \rho_0^{-\gamma}=P_p \rho_p^{-\gamma} \ , \label{eq:polytrope}
\end{equation}
where $\gamma$ is the specific heat ratio. We assume $\gamma=5/3$ (monoatomic ideal gas).

There are several discontinuities in the reconnection outflow, i.e., X-shaped slow shock, reverse fast shock, contact discontinuity, and forward V-shaped slow shock (see figure \ref{fig:structure-sch}). We set jump conditions for both sides of each discontinuity: First we consider the X-shaped slow shock Rankine-Hugoniot (R-H) jump conditions. The pressure-jump is 
\begin{eqnarray}
\frac{P_1}{P_p}=&1+\frac{\gamma}{c_{sp}^2}(-\sin \theta v_{xp} + \cos \theta v_{yp})^2 (X-1) \times
\{
\frac{1}{X}-(\cos \theta B_{xp}+\sin \theta B_{yp})^2/2 \nonumber \\
&\times [-2 V_{Apx}^2 X+(-\sin \theta v_{xp} + \cos \theta v_{yp})^2 (X+1)]/[((-\sin \theta v_{xp} + \cos \theta v_{yp})^2-X V_{Apx}^2)^2 \mu \rho_p]
\} \ ,\label{eq:p-p-jump}
\end{eqnarray}

where
$$c_{sp}=\sqrt{\gamma P_p/\rho_p} \ ,$$ 
$$V_{Apx}=\sqrt{(-\sin \theta B_{xp}+\cos \theta B_{yp})^2/(\mu \rho_p)} \ ,$$
$\mu$ is the magnetic permeability of vacuum and $X$ is the compression ratio. \\

Then the density-jump is
\begin{equation}
\frac{\rho_1}{\rho_p}=X \ .\label{eq:p-d-jump}
\end{equation}

The velocity (parallel) jump is
\begin{equation}
\frac{\cos \theta (v_1-x_y)-v_0}{\sin \theta v_{yp}-v_0}=\frac{(-\sin \theta v_{xp} + \cos \theta v_{yp})^2-V_{Apx}^2}{(-\sin \theta v_{xp} + \cos \theta v_{yp}^2-X V_{Apx}^2)} \ ,\label{eq:p-vpara-jump}
\end{equation}

where $v_0=(v_{yp} B_{xp}-B_{yp} v_{xp})/(B_{xp} \sin \theta-B_{yp} \cos \theta)$ is the shift speed of the de Hoffmann-Teller coordinate.

The velocity (perpendicular) jump is
\begin{equation}
\frac{-\sin \theta (v_1-x_y)}{-\sin \theta v_{xp} + \cos \theta y_{yp}}=\frac{1}{X} \ .\label{eq:p-vperp-jump}
\end{equation}

The magnetic field (parallel) jump is
\begin{equation}
\frac{\sin \theta B_1}{\cos \theta B_{xp}+\sin \theta B_{yp}}=\frac{[(-\sin \theta v_{xp} + \cos \theta v_{yp})^2-V_{Apx}^2]X}{(-\sin \theta v_{xp} + \cos \theta v_{yp})^2-X V_{Apx}^2} \ .\label{eq:p-bpara-jump}
\end{equation}

The magnetic field (perpendicular) jump is
\begin{equation}
\frac{\cos \theta B_1}{-\sin \theta B_{xp}+\cos \theta B_{yp}}=1 \ .\label{eq:p-bperp-jump}
\end{equation}

The compression ratio $X$ is defined by the following equation (3rd order algebraic equation for $X$),
\begin{eqnarray}
&\{
(-\sin \theta B_{xp}+\cos \theta B_{yp})^2 [(\cos \theta B_{xp}+\sin \theta B_{yp})^2 (\gamma - 1) \rho_p (-\sin \theta v_{xp} + \cos \theta v_{yp})^2 \nonumber \\
&+(-\sin \theta B_{yp}+\cos \theta B_{yp})^2 (2 \gamma P_p-\rho_p (-\sin \theta v_{xp} + \cos \theta v_{yp})^2+\gamma \rho_p (-\sin \theta v_{xp} + \cos \theta v_{yp})^2)]
\} X^3 \nonumber \\
&+
\{
-\rho_p (-\sin \theta v_{xp} + \cos \theta v_{yp})^2 
[
(-\sin \theta B_{xp}+\cos \theta B_{yp})^4 (\gamma-1) \nonumber \\
&+
(\cos \theta B_{xp}+\sin \theta B_{yp})^2 (\gamma-2) \mu \rho_p (-\sin \theta v_{xp} + \cos \theta v_{yp})^2 \nonumber \\
&+
(-\sin \theta B_{xp}+\cos \theta B_{yp})^2
(
(\cos \theta B_{xp}+\sin \theta B_{yp})^2 (\gamma+1) \nonumber \\
&+
2 \mu (2 \gamma P_p-\rho_p (-\sin \theta v_{xp} + \cos \theta) v_{yp})^2+\gamma \rho_p (-\sin \theta v_{xp} + \cos \theta v_{yp})^2
)
]
\} X^2 \nonumber \\
&+
\{
\mu \rho_p (-\sin \theta v_{xp} + \cos \theta v_{yp}^4
[
(\cos \theta B_{xp}+\sin \theta B_{yp})^2 \gamma+
2(-\sin \theta B_{xp}+\cos \theta B_{yp})^2 (\gamma+1)+
\mu \nonumber \\
&(
2 \gamma P_p-\rho_p (-\sin \theta v_{xp} + \cos \theta v_{yp})^2+\gamma \rho_p (-\sin \theta v_{xp} + \cos \theta v_{yp})^2
)
]
\} X \nonumber \\
&+
\{
-(\gamma+1) \mu^2 \rho_p^3 (-\sin \theta v_{xp} + \cos \theta v_{yp})^6
\}=0 \ . \nonumber
\end{eqnarray} 
This equation has three roots. We must choose a real positive root larger than unity.

\noindent
The reverse-fast shock R-H conditions are as follows. The pressure-jump is
\begin{equation}
\frac{P_2}{P_1}=\zeta_f \ ,\label{eq:f-p-jump}
\end{equation}

where 
$$\zeta_f=\gamma M_{1f}^2 (1-1/\xi_f)+(1-\xi_f^2)/\beta_1+1$$ 
with 
$$\xi_f=(-l + \sqrt{l^2 + 2/\beta_1 (2 - \gamma) (\gamma + 1) \gamma M_{1f}^2})/(2/\beta_1 (2 - \gamma)) \ ,$$
$$l = \gamma (1/\beta_1 + 1) + (\gamma - 1) \gamma M_{1f}^2/2 \ ,$$
$$\beta_1 = P_1/(B_1^2/(2 \mu)) \ ,$$
$$M_{1f} = (v_1 - x_f V_A)/c_f \ ,$$
$$c_f = \sqrt{\gamma P_1/\rho_1}$$
and 
$$V_A = B_0/\sqrt{\mu \rho_0}\ .$$

The density-jump is
\begin{equation}
\frac{\rho_2}{\rho_1}=\xi_f \ .\label{eq:f-d-jump}
\end{equation}

The velocity jump is
\begin{equation}
\frac{v_1-x_f}{v_2-x_f}=\xi_f \ .\label{eq:f-v-jump}
\end{equation}

The magnetic field jump is
\begin{equation}
\frac{B_2}{B_1}=\xi_f \ .\label{eq:f-b-jump}
\end{equation}

We must impose local magnetic flux conservation on both sides of region p and region 1. The magnetic flux conservation at X-point gives
\begin{equation}
-v_{xp} B_{yp} + v_{yp} B_{xp}+(v_1-x_y) B_1=0 \ .\label{eq:bflux-cons}
\end{equation}

Force balance at contact discontinuity is
\begin{equation}
P_2+\frac{B_2^2}{2 \mu}=P_3+\frac{B_{x3}^2+B_{y3}^2}{2 \mu} \ .\label{eq:f-balance}
\end{equation}

The V-shaped forward slow shock R-H conditions are as follows. The pressure-jump is
\begin{equation}
P_3=P_0 \left\{
1+\frac{\gamma}{c_{s0}^2} (x_s \sin \phi)^2 (X_h-1)
\times \left[
\frac{1}{X_h}-\frac{(B_0 \cos \phi)^2}{2} \frac{-2 V_{A0x}^2 X_h+(x_s \sin \phi)^2 (X_h+1)}{[(x_s \sin \phi)^2-X_h V_{A0x}^2]^2 \mu \rho_0}
\right]
\right\} \ ,\label{eq:s-p-jump}
\end{equation}
where
$c_{s0}=\sqrt{\gamma P_0/\rho_0}$, $V_{A0x}=\sqrt{(-\sin \phi B_0)^2/(\mu \rho_0)}$ and $X_h$ is the compression ratio. The density-jump is
\begin{equation}
\frac{\rho_3}{\rho_0}=X_h \ .\label{eq:s-d-jump}
\end{equation}

The velocity (parallel) jump gives
\begin{equation}
\frac{\cos \phi (v_{x3}-x_s V_{A0})+\sin \phi v_{y3}}{-V_{A0} x_s \cos \phi}
=\frac{v_{m0x}^2-V_{A0x}^2}{v_{m0x}^2-X_h V_{A0x}^2} \ ,\label{eq:s-vpara-jump}
\end{equation}
where $v_{m0x}=x_s \sin \phi$ and $V_{A0}=B_0/\sqrt{\mu \rho_0}$. The velocity (perpendicular) jump gives
\begin{equation}
\frac{-\sin\phi (v_{x3}-x_s V_{A0})+\cos \phi v_{y3}}{V_{A0} x_s \sin \phi}=\frac{1}{X_h} \ .\label{eq:s-vperp-jump}
\end{equation}

The magnetic field (parallel) jump is
\begin{equation}
\frac{\cos \phi B_{x3}+\sin \phi B_{y3}}{B_0 \cos \phi}=\frac{(v_{m0x}^2-V_{A0x}^2) X_h}{v_{m0x}^2-X_h V_{A0x}^2} \ .\label{eq:s-bpara-jump}
\end{equation}

The magnetic field (perpendicular) jump is
\begin{equation}
\frac{-\sin \phi B_{x3}+\cos \phi B_{y3}}{-B_0 \sin \phi}=1 \ .\label{eq:s-bperp-jump}
\end{equation}

The compression ratio $X_h$ is a solution of the following third order algebraic equation:
\begin{eqnarray}
&\{
(B_0 \sin \phi)^2 
[
(B_0 \sin \phi)^2 (\gamma-1) \rho_0 (x_s \sin \phi)^2+
(B_0 \sin \phi)^2 
(
2 \gamma P_0-\rho_0 (x_s \sin \phi)^2+\gamma \rho_0 (x_s \sin \phi)^2
)
]
\} X_h^3 \nonumber \\
&+
\{
-\rho_0 (x_s \sin \phi)^2
[
(B_0 \sin \phi)^4 (\gamma+1)+(B_0 \cos \phi)^2 (\gamma-2) \mu \rho_0 (x_s \sin \phi)^2 \nonumber \\
&+(B_0 \sin \phi)^2 
(
(B_0 \cos \phi)^2 (\gamma+1)+(2 \mu (2 \gamma P_0-\rho_0 (x_s \sin \phi)^2+\gamma \rho_0 (x_s \sin \phi)^2))
)
]
\} X_h^2 \nonumber \\
&+
\{
\mu \rho_0^2 (x_s \sin \phi)^4 
(
(B_0 \cos \phi)^2 \gamma + 2(B_0 \sin \phi)^2 (\gamma+1)+
\mu (2 \gamma P_0 - \rho_0 (x_s \sin \phi)^2+\gamma \rho_0 (x_s \sin \phi)^2)
)
\} X_h \nonumber \\
&+
\{
-(\gamma+1) \mu^2 \rho_0^3 (x_s \sin \phi)^6
\}=0\ . \nonumber
\end{eqnarray}

We must choose a real positive root larger than unity. 

The tip of the reconnection outflow touches the FRWF (see figure \ref{fig:structure-sch}), hence $A_1'=0$ at this point. This condition reduces to the following boundary condition at the tip of the outflow:  

\begin{equation}
-B_{y3} (1-x_c) + B_{x3} [(1-x_s) \tan \phi - (x_c-x_y) \tan \theta]-B_0 (1-x_s) \tan \phi =0 \label{eq:bc-tip}
\end{equation}

From magnetic flux conservation, the injected magnetic flux must be redistributed in the reconnection jet. This leads to the following equation for magnetic flux conservation:  

\begin{equation}
-B_{yp} v_{xp}+v_{yp} B_{xp}+[(x_f-x_y) B_1+ (x_c-x_f) B_2]=0 \label{eq:bflux-cons2}
\end{equation}

We assume the following trivial relations:\\

\noindent
Definition from the contact discontinuity
$$
x_c=v_2=v_{x3} 
$$

We can solve [\ref {eq:polytrope}] for $P_p$, [\ref {eq:p-bperp-jump}] for $B_1$, [\ref {eq:f-p-jump}] for $P_2$, [\ref {eq:f-d-jump}] for $\rho_2$, [\ref {eq:f-b-jump}] for $B_2$, [\ref {eq:bflux-cons}] for $v_{yp}$, [\ref {eq:s-p-jump}] for $P_3$, [\ref {eq:s-d-jump}] for $\rho_3$, [\ref {eq:s-vpara-jump}] and [\ref {eq:s-vperp-jump}] for $v_{x3}$ and $v_{y3}$, [\ref {eq:s-bpara-jump}] and [\ref {eq:s-bperp-jump}] for $B_{x3}$ and $B_{y3}$ by hand, then substitute them into other nine equations ([\ref {eq:frozen-in}], [\ref {eq:p-p-jump}], [\ref {eq:p-d-jump}], [\ref {eq:p-vpara-jump}], [\ref {eq:p-vperp-jump}], [\ref {eq:p-bpara-jump}], [\ref {eq:f-v-jump}], [\ref {eq:f-balance}], [\ref {eq:bc-tip}] and [\ref {eq:bflux-cons2}]) for the following unknowns: $\rho_p$, $v_{xp}$, $B_{xp}$, $B_{yp}$, $\theta$, $P_1$, $\rho_1$, $x_f$, $\phi$ and $x_s$. The only parameter included in this problem is the plasma-$\beta$ value at the asymptotic region. By using a Newton-Raphson routine, with an initial guess for these unknowns, we obtain converged solutions. The procedure to obtain the series of converged solutions is discussed in section \ref{sec:b-eqs} in detail.

As $|v_{yp}|$ decreases, the strength of reverse fast shock reduces, and then the pressure jump $\zeta_f$ and density jump $\xi_f$ simultaneously become unity at the critical value of $|v_{yp}|$. At that point, the reverse fast shock vanishes. We convert the coupled equations to@another set compatible to the situation with no fast shock, i.e., several equations reform to the following equations: 

\noindent
[\ref {eq:f-p-jump}] is replaced by\\
$$P_2=P_1 \ ,$$ 

\noindent
[\ref {eq:f-d-jump}] is replaced by\\
$$\rho_2=\rho_1 \ ,$$ 

\noindent
[\ref {eq:f-v-jump}] is replaced by\\
$$v_1=v_2 \ ,$$

\noindent
[\ref {eq:f-b-jump}] is replaced to by\\
$$B_1=B_2 \ ,$$ 

and [\ref {eq:bflux-cons2}] becomes to be equivalent to [\ref {eq:bflux-cons}] and removed from the set of coupled equations. The number of coupled equations reduces to 21. This is consistent with that the locus $x_f$ of the fast shock is no longer included in the set of unknowns and total number of unknowns becomes 21.

\appendix

\figcaption[]{Schematic figure of the reconnection outflow. Several discontinuities form in the outflow. We consider a quasi-one dimensional shock-tube problem along the $x-$axis and $x=x_y$. 
\label{fig:structure-sch}
}

\figcaption[]{Schematic picture of the inflow structure observed in the ``fixed'' (initial plasma-comoving) coordinate. We assume the vertical inflow in order to keep consistency with our linearized analysis via Grad-Shafranov approach (Nitta et al. 2002). Note that the Alfv\'{e}n Mach number of  this inflow does NOT mean the reconnection rate. 
\label{fig:fixed-flow-sch}
}

\figcaption[]{Schematic picture of the inflow structure observed in the ``Zoom-out'' (reconnection point-comoving) coordinate. An apparent convergence (the horizontal component) of the inflow takes place. This horizontal velocity component increases as the magnetic Reynolds number $R_{em}^*$ increases. The resultant oblique inflow tends to be parallel to the magnetic field lines as $R_{em}^*$ increases. Thus the reconnection rate decreases as $R_{em}^*$ increases. 
\label{fig:comoving-flow-sch}
}

\figcaption[]{Magnetic Reynolds number dependence of the reconnection rate $R^*$. This is the case for $\beta_0=0.01$. The reconnection rate decreases as $\propto 1/R_{em}^*$ in very large magnetic Reynolds number. 
\label{fig:rec-rate}
}

\figcaption[]{Velocity and magnetic field at the region p versus magnetic Reynolds number $R_{em}^*$ in the zoom-out (reconnection point-comoving) frame. This is the case for $\beta_0=0.01$. These quantities are essential to determine the reconnection rate $R^*$. As $R_{em}^*$ increases, first, the transverse component $B_{xp}$ of the magnetic field quickly increases. Then, $v_{xp}$ drastically developed. The negative value of $v_{xp}$ denotes an apparent converging inflow toward the Petschek-like slow shock. Note $|v_{xp}|$ does not exceed unity even in the range of very large magnetic Reynolds number. 
\label{fig:quantities}
}

\figcaption[]{Direction of velocity and magnetic field at the region p in the zoom-out (reconnection point-comoving) frame. This is the case for $\beta_0=0.01$. The circular dot and square denote $v_{yp}/v_{xp}$ (tangent of the inclination angle of the inflow velocity) and $B_{yp}/B_{xp}$ (tangent of the inclination angle of the inflow magnetic field), respectively. This graph  clearly shows that the inflow velocity $\mbox{\boldmath v}_p$ and the inflow magnetic field $\mbox{\boldmath B}_p$ tend to be parallel to each other when the magnetic Reynolds number $R_{em}^*$ increases. This causes the decrement of the reconnection rate $R^* \equiv \mbox{\boldmath v}_p \times \mbox{\boldmath B}_p$. We should note that the information of the inflow Alfv\'{e}n Mach number ($v_{yp}$) perpendicular to the current sheet is not sufficient to determine the reconnection rate. 
\label{fig:tilt}
}

\figcaption[]{The magnetic field line configuration (Petschek type) in the inflow region for the case $R_{em}^*=15.63$ and $\beta_0=0.01$ derived from linearized G-S approach. The single reconnection point forms at $x=0$. The entire structure is very similar to that of the Petschek model. We call this the single X-point solution. The reconnection rate in this case is the maximum value $R^*=0.0504$. Note that only the region inside the FRWF ($x^2+y^2<1$) is solved in this figure. The region outside the FRWF ($x^2+y^2>1$) is in the initial uniform equilibrium. 
\label{fig:single-X}
}

\figcaption[]{The magnetic field line configuration (X-O-X type) in the inflow region for the case $R_{em}^*=16.88$ and $\beta_0=0.01$. The double reconnection points are formed at $x=\pm 0.048$. A magnetic island forms between them. This is the X-O-X type solution. The reconnection rate in this case is $R^*=0.0489$. Note that only the region inside the FRWF ($x^2+y^2<1$) is solved in this figure. The region outside the FRWF ($x^2+y^2>1$) is in the initial uniform equilibrium. 
\label{fig:X-O-X}
}

\figcaption[]{The magnetic field line configuration (double Y-point type) in the inflow region for the case $R_{em}^*=139.8$ and $\beta_0=0.01$. A magnetic island forms between the double reconnection points at $x=\pm 0.75$. The island is extremely elongated and looks like a current sheet. This is a new type of slow regime. We call it the double Y-point solution. The reconnection rate in this case is $R^*=0.00713$. Note that only the region inside the FRWF ($x^2+y^2<1$) is solved in this figure. The region outside the FRWF ($x^2+y^2>1$) is in the initial uniform equilibrium. Although this solution looks like the Sweet-Parker solution, there are critical differences between them (see section \ref{sec:dis-transition}). 
\label{fig:double-Y}
}

\figcaption[]{The schematic picture of the reducible region. The plasma in the region below the separatrix stream line (SSL) can enter the Petschek-like slow shock (SS) elongated from the reconnection point. Such plasma can release its magnetic energy. By noting the self-similar expansion of the system, the magnetic energy in the region below the dashed line having the inclination $\theta_R$ from the $x$-axis is reducible by single reconnection. We approximate the separatrix stream line to be straight and that the slow shock completely converts the magnetic energy for simplicity in order to estimate the reducible fraction.
\label{fig:reducible-sch}
}

\figcaption[]{The reducible fraction of the magnetic energy of the initial current sheet system is shown as a function of the magnetic Reynolds number under the approximation that the separatrix stream line is straight. This is the case for $\beta_0=0.01$. When the magnetic Reynolds number is small enough ($R_{em}^*=15.63$), the reconnection can completely convert the magnetic energy ($f=1$). As the magnetic Reynolds number increases, the reducible fraction drastically decreases. Finally it tends to $f \propto {R_{em}^*}^{-1/2}$. This means that the released energy cannot be determined even if we estimate the magnetic energy of the current sheet system until we know the magnetic Reynolds number. 
\label{fig:reducible}
}

\figcaption[]{The locus of the reconnection point ($x=x_y$) as a function of the magnetic Reynolds number $R_{em}^*$. This is the case for $\beta_0=0.01$. We assumed vertical inflow in the fixed coordinate (thus $x_y=-v_{xp}$). When the resistivity is large enough (the magnetic Reynolds number $R_{em}^*=15.63$), $x_y=0$ (Petschek-like solution). As $R_{em}^*$ increases, $x_y$ also increases. In the limit of large $R_{em}^*$, $x_y$ approaches to unity. 
\label{fig:xy}
} 

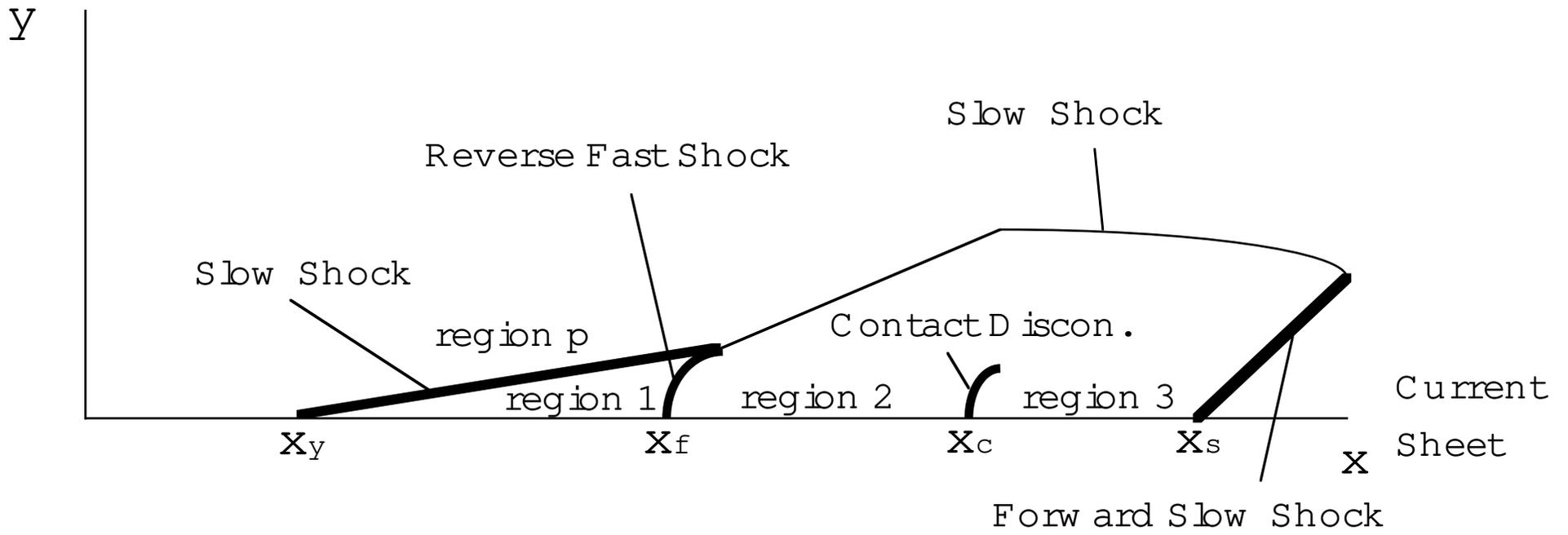
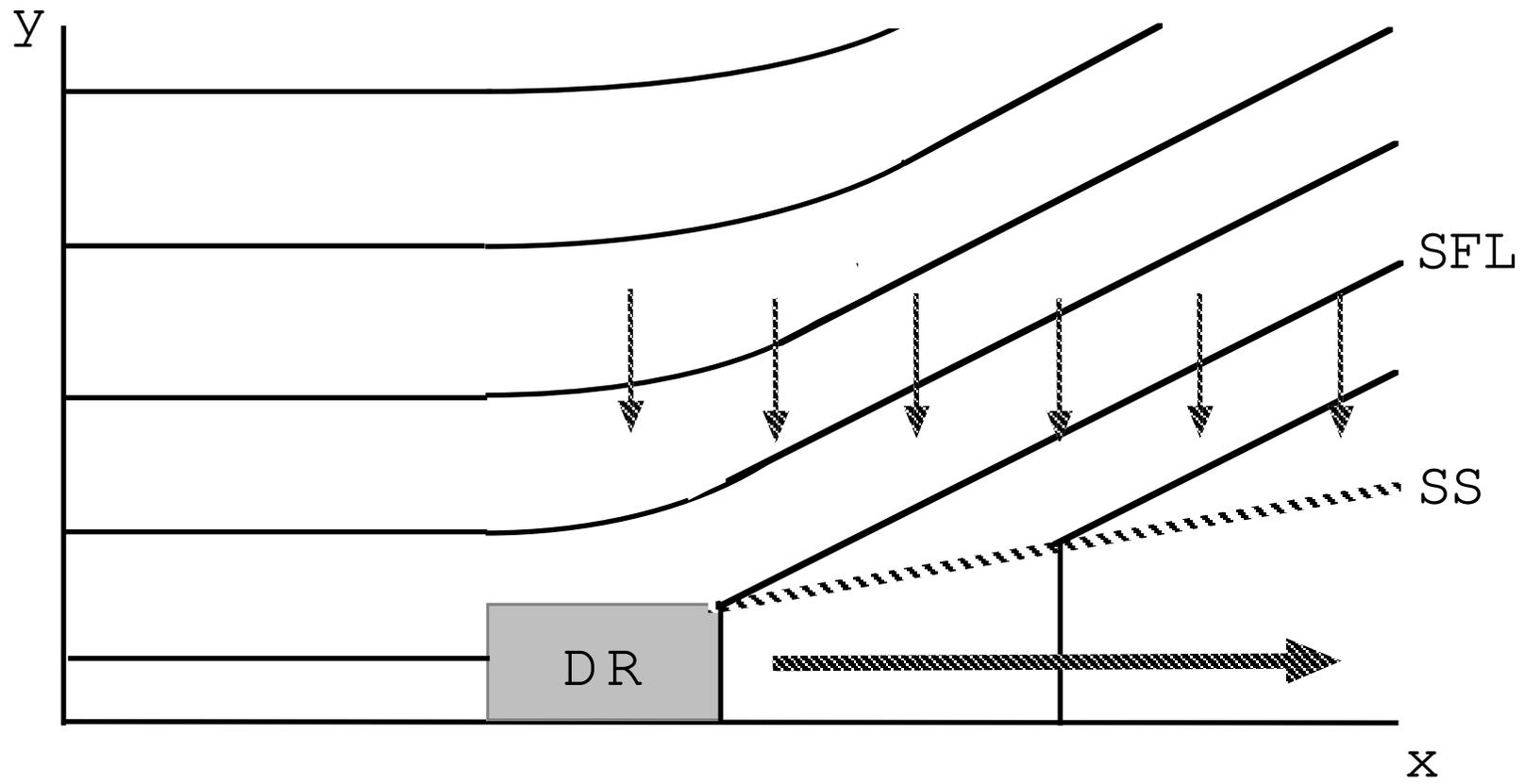
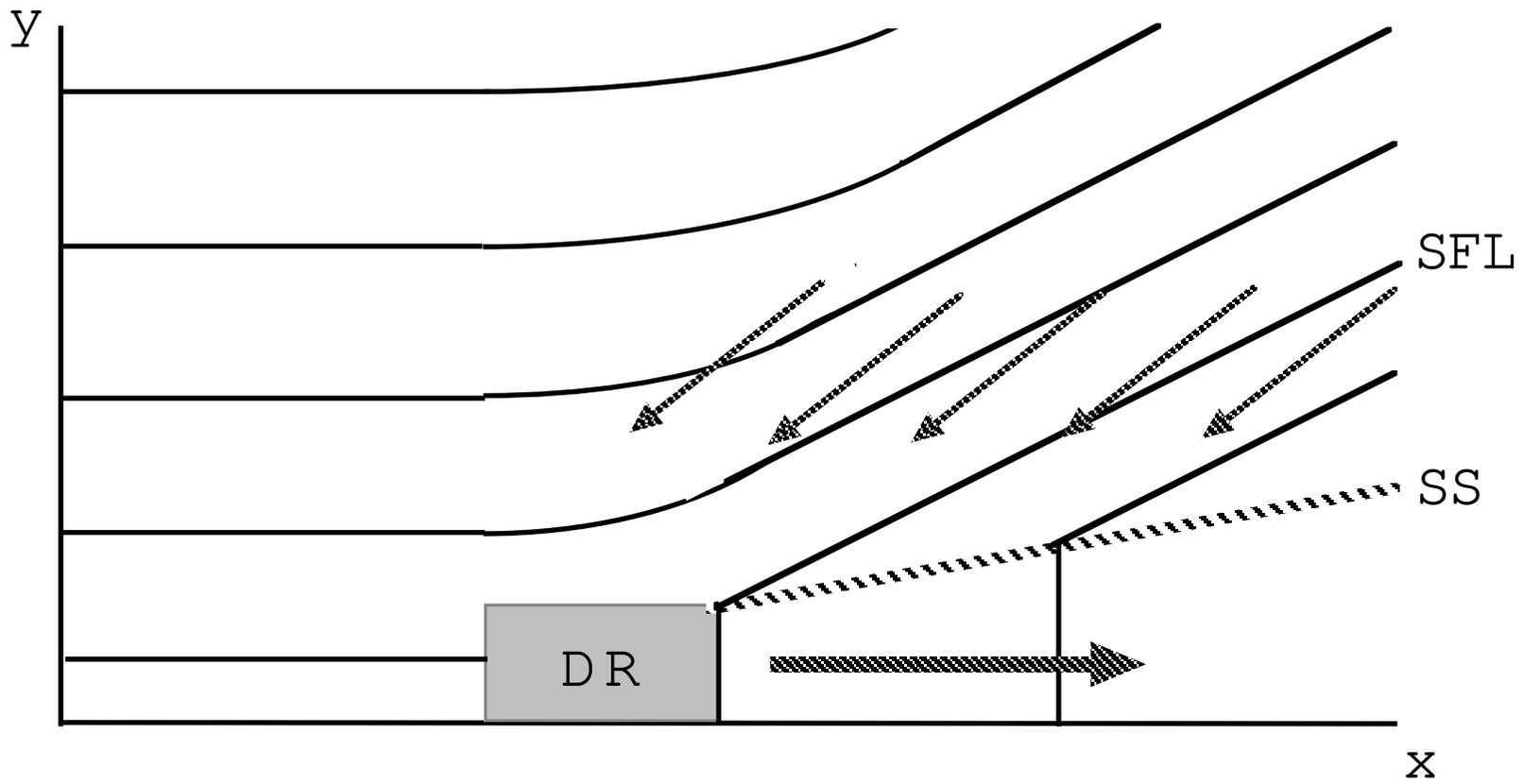
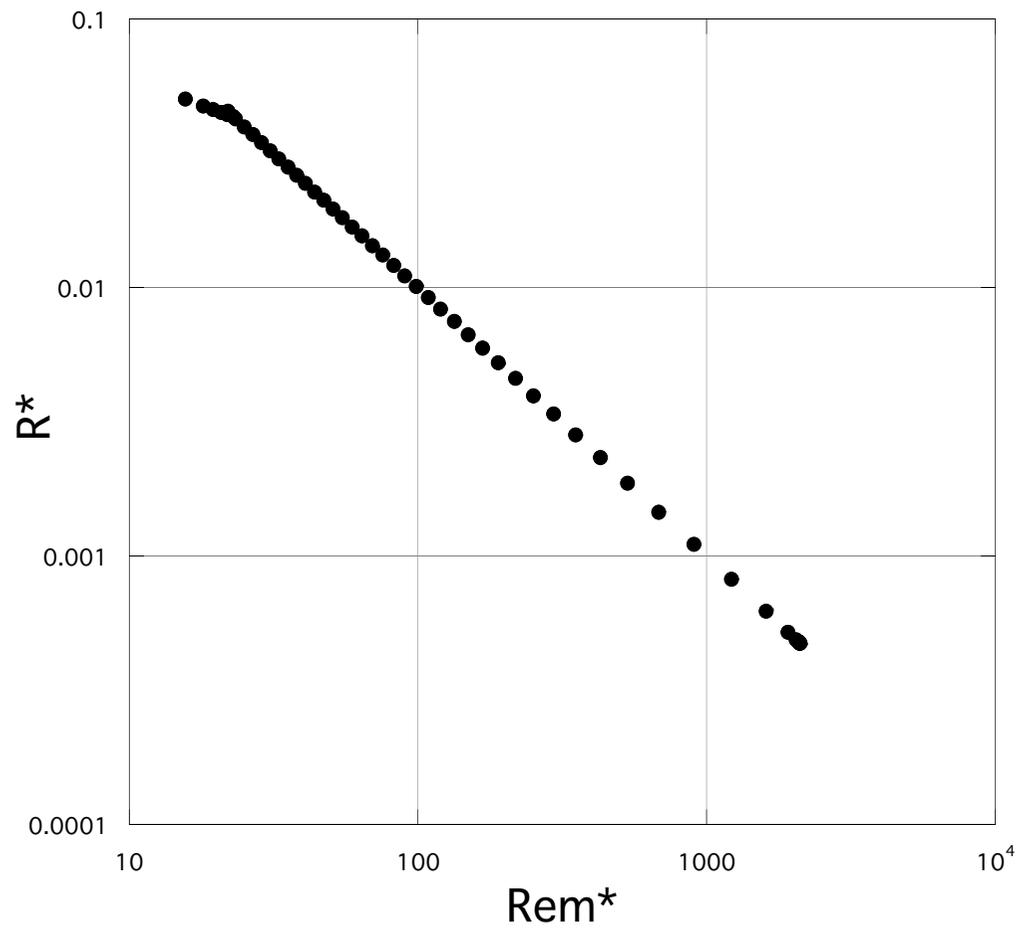
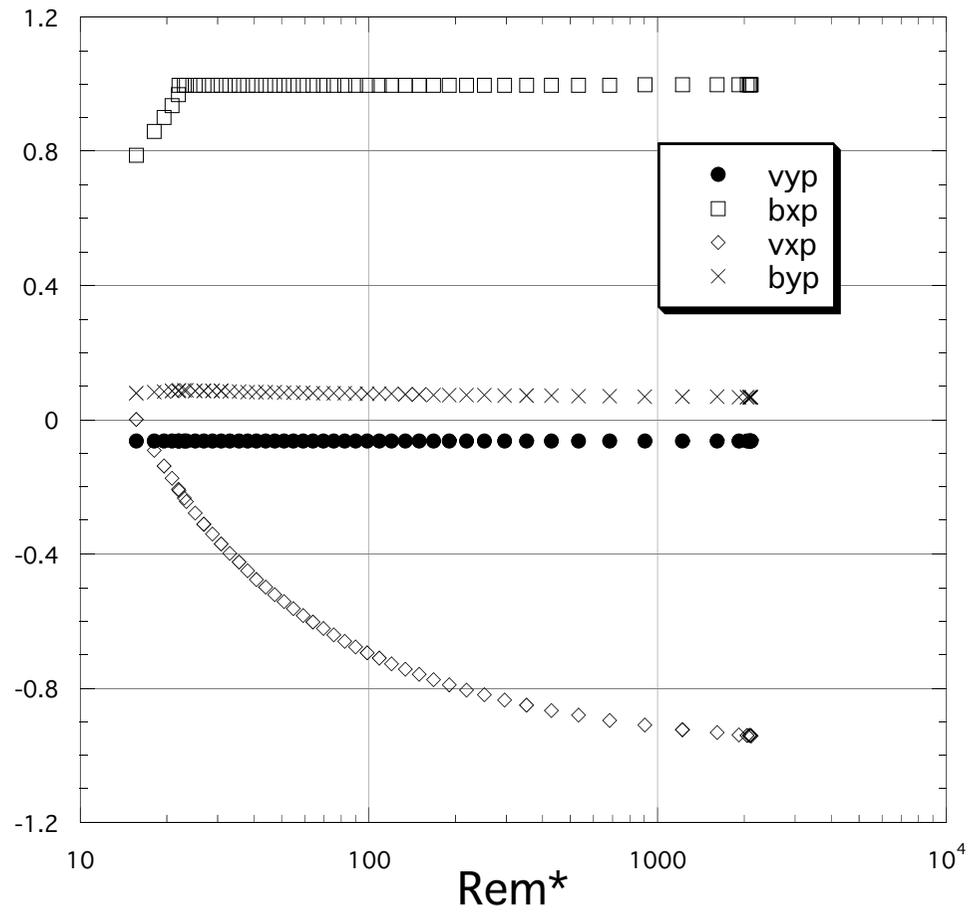
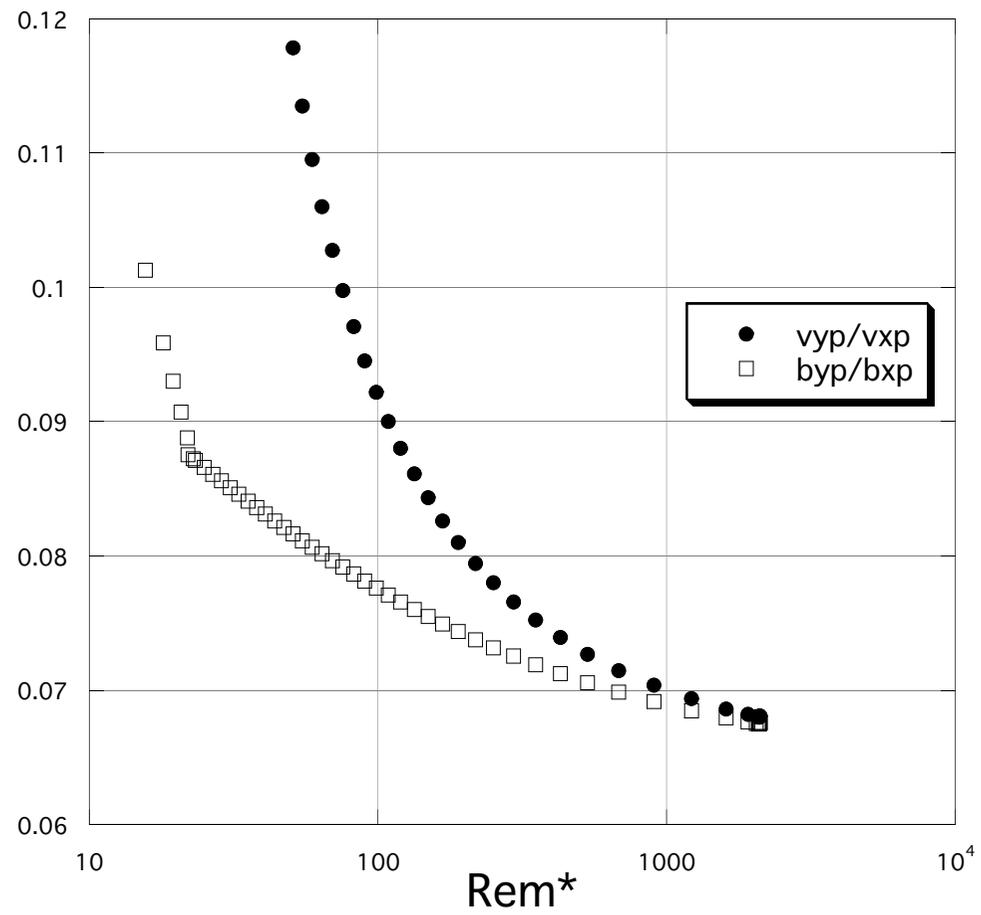
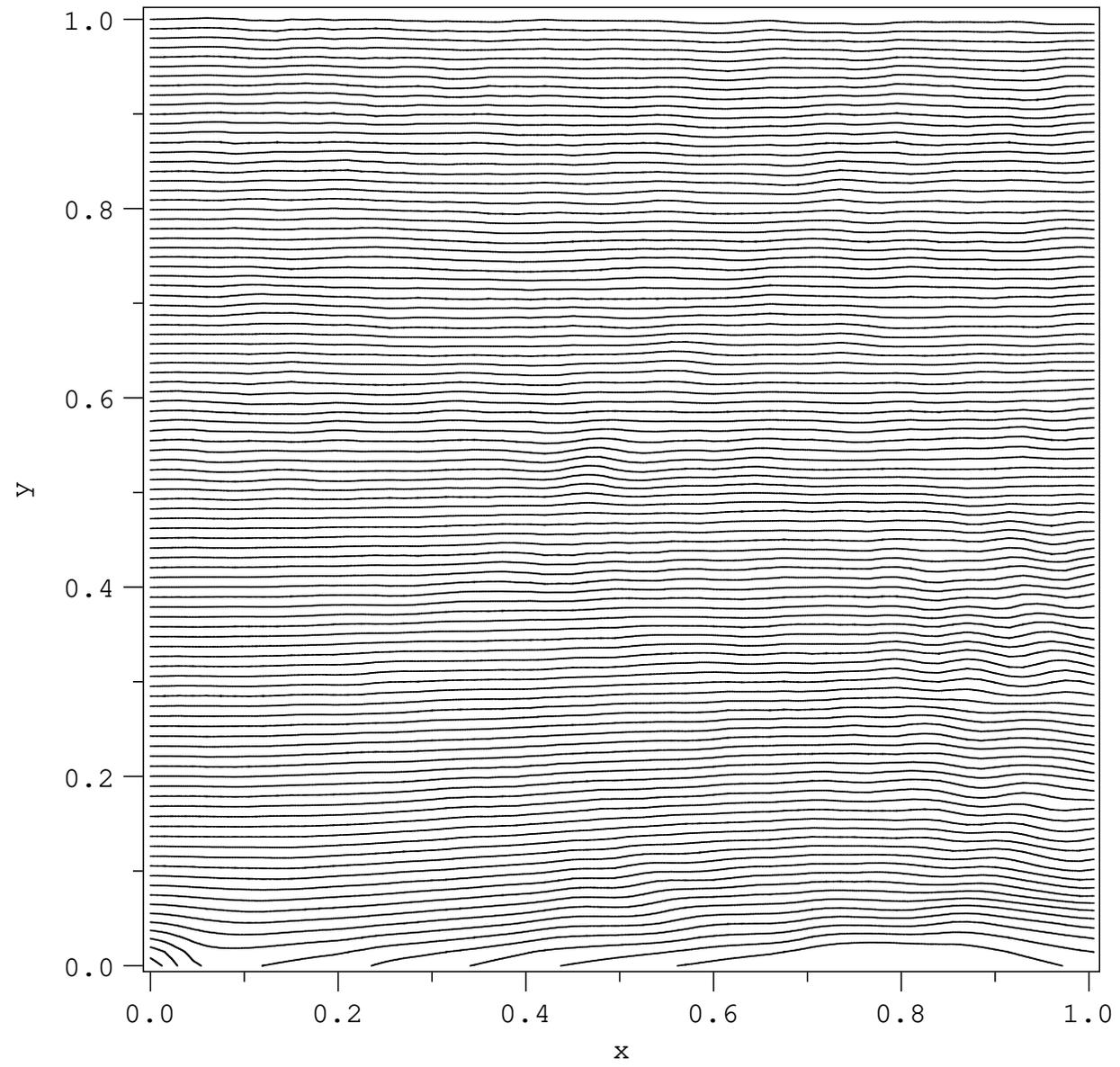
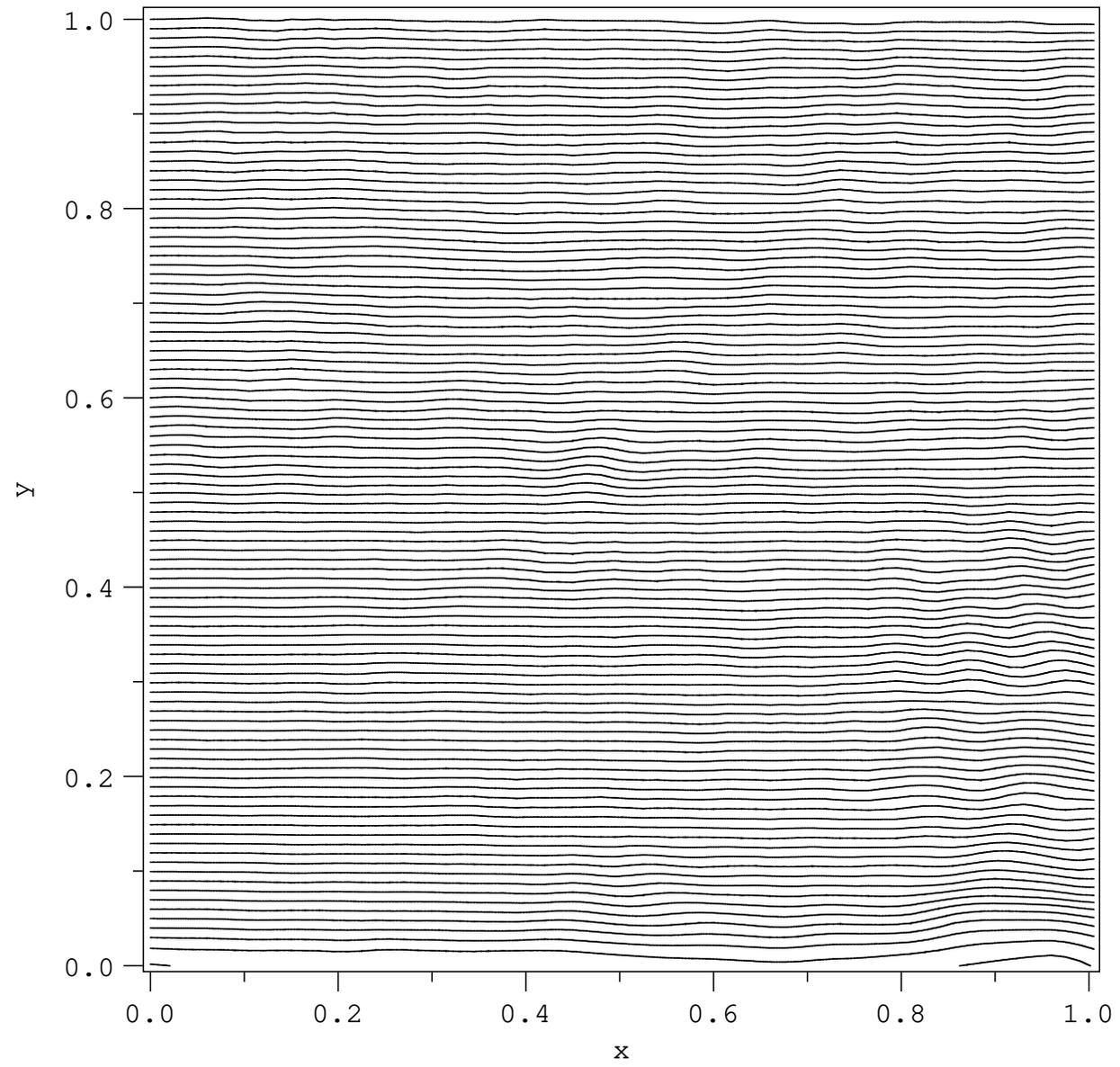
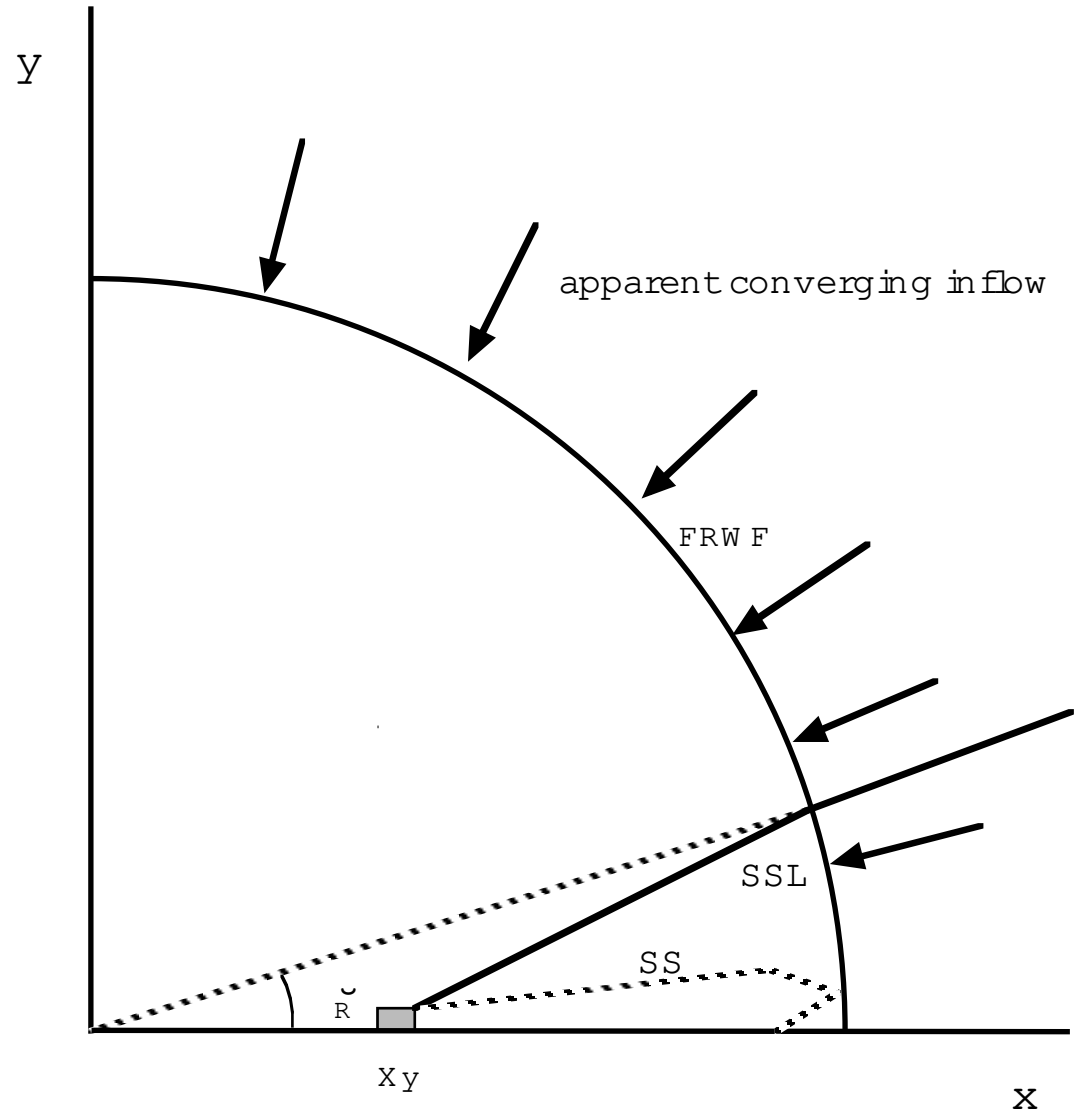
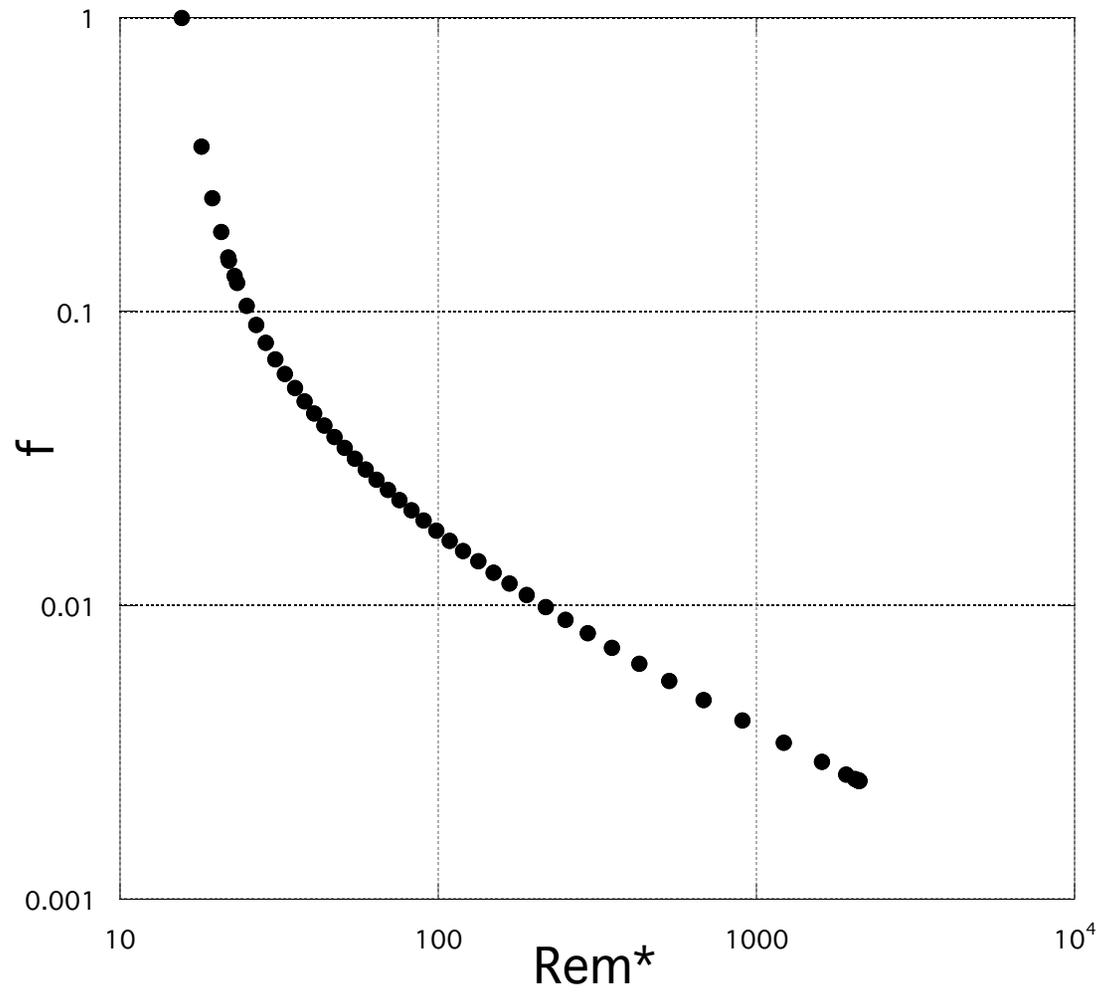
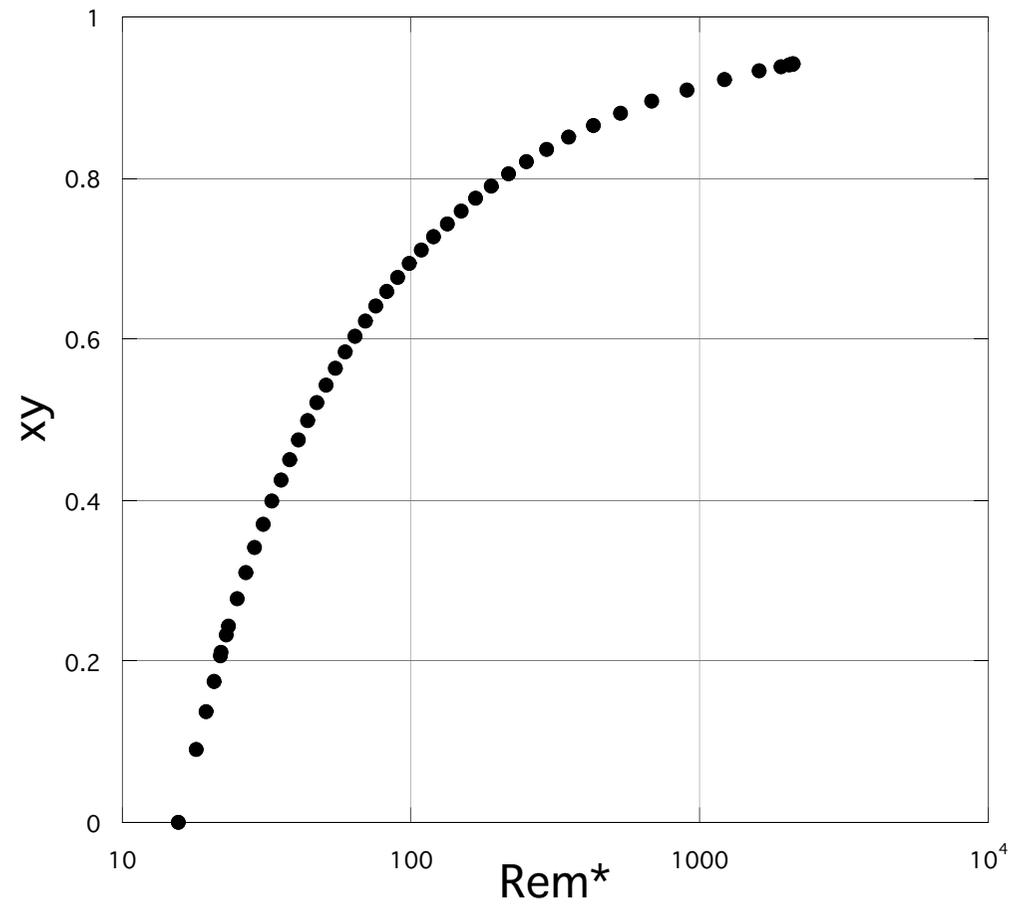
\end{document}